\documentclass[a4paper,fleqn]{cas-sc}

\usepackage[authoryear]{natbib}

\def\tsc#1{\csdef{#1}{\textsc{\lowercase{#1}}\xspace}}
\tsc{WGM}
\tsc{QE}

\PassOptionsToPackage{table,xcdraw,dvipsnames, usenames}{xcolor}
\usepackage{graphicx}
\usepackage{subcaption}
\usepackage{url}
\usepackage{enumitem}
\usepackage{tcolorbox}
\usepackage{multirow}

\usepackage{mathtools}
\DeclarePairedDelimiter{\evdel}{\langle}{\rangle}
\newcommand{\ev}{\operatorname{}\evdel}
\usepackage{array,multirow}
\usepackage{amsmath}
\usepackage{adjustbox}
\usepackage{booktabs}
\usepackage{soul}
\usepackage[english]{babel}
\usepackage{anyfontsize}
\usepackage{algorithm}
\usepackage{float}
\usepackage[]{xcolor}

\usepackage[normalem]{ulem}
\useunder{\uline}{\ul}{}

\graphicspath{{figures/}} %

\usepackage{todonotes}
\newcommand{\sectopic}[1]{\vspace{0em}\par\noindent{\textit{\bfseries #1}}}

\newcommand{\rqdef}[1]{\vspace{1em}\par\noindent{\bfseries #1}}

\usepackage{varwidth}
\usepackage{ragged2e}
\usetikzlibrary{calc}

\newcounter{tipcounter}
\usepackage{tikz}

\def\blue#1           {\textcolor{blue} {#1} }

\newcommand{\gensys}{\textsf{REIT}}
\newcommand{\sys}{\textsf{RoREIT}}
\newcommand{\tts}{\textsf{VoREIT}}

\DeclareMathAlphabet{\mathit}{\encodingdefault}{\familydefault}{m}{it}

\begin{document}
\let\WriteBookmarks\relax
\def\floatpagepagefraction{1}
\def\textpagefraction{.001}

\shorttitle{Exploring Emerging Technologies for Requirements Elicitation Interview Training}    

\shortauthors{Görer, Binnur, et al.}  

\title [mode = title]{Exploring Emerging Technologies for Requirements Elicitation Interview Training: Empirical Assessment of Robotic and Virtual Tutors}  




%

\author[1]{Binnur Görer}[orcid=0000-0001-9153-9244]

\cormark[1]


\ead{binnur.gorer@boun.edu.tr}


\credit{Conceptualization, Method, Software, Validation, Investigation, Data Curation, Writing}

\affiliation[1]{organization={Boğaziçi University},
            city={Istanbul},
            country={Türkiye}}

\author[1]{Fatma Başak Aydemir}[orcid=0000-0003-3833-3997]


\ead{basak.aydemir@boun.edu.tr}


\credit{Conceptualization, Method, Writing, Supervision}

\cortext[1]{Corresponding author}



\begin{abstract}
Requirements elicitation interviews are a widely adopted technique, where the interview success heavily depends on the interviewer's preparedness and communication skills. Students can enhance these skills through practice interviews. 
However, organizing practice interviews for many students presents scalability challenges, given the time and effort required to involve stakeholders in each session.
To address this, we propose \gensys{}, an extensible architecture for Requirements Elicitation Interview Training system based on emerging educational technologies. \gensys{} has components to support both the interview phase, wherein students act as interviewers while the system assumes the role of an interviewee, and the feedback phase, during which the system assesses students' performance and offers contextual and behavioral feedback to enhance their interviewing skills.  
We demonstrate the applicability of \gensys{} through two implementations: \sys{} with a physical robotic agent and \tts{} with a virtual voice-only agent. We empirically evaluated both instances with a group of graduate students. The participants appreciated both systems. They demonstrated higher learning gain when trained with \sys{}, but they found \tts{} more engaging and easier to use. These findings indicate that each system has distinct benefits and drawbacks, suggesting that \gensys{} can be realized for various educational settings based on preferences and available resources.

\end{abstract}


\begin{highlights}
    \item We propose the extensible architecture \gensys{} for requirements elicitation training systems and demonstrate its applicability by implementing two instances of it with different agent structures and interaction modalities.
    \item We empirically evaluate the two interview training systems -- \sys{} with an embodied robotic agent and \tts{} with a virtual voice agent -- with the students of a graduate level requirements engineering course.
    \item The participants rated \tts{} more favorably for the ease of use and engagement while \sys{} yielded a notably higher learning gain compared to \tts{}.
    \item Our findings indicate that each system has its own distinct advantages and weaknesses. Software engineering educators can customize \gensys{} based on their needs and available resources. 
    \item We publicly share our system implementation and study materials in~\citet{binnur_gorer_2023_7861906}.
\end{highlights}


\maketitle

\section{Introduction}\label{sec:introduction}

Requirements elicitation is a set of activities to gather stakeholders' needs and desires for a system-to-be~\citep{zowghi2005requirements} and is crucial for the success of software projects. Software developers can reduce project risks, improve team communication, and deliver high-quality software products that meet customer needs and desires by investing time and effort in requirements elicitation~\citep{van2009requirements}. Among various elicitation techniques, interviews are the most popular and effective~\citep{davis2006effectiveness}.

Requirements elicitation interviews require a combination of theoretical knowledge of interview techniques and soft skills such as interview management, behavioral control, and confidence~\citep{hadar2014role}. It is essential to practice in a real interview setting to improve these soft skills and overcome interview nervousness~\citep{andrews2006student,powell2021shake}. A popular teaching strategy for requirements elicitation interviews is role-playing, which allows students to improve their abilities while playing the roles of stakeholders and requirements engineers. However, the excessive human effort required to plan and oversee these activities often makes them impractical in a regular classroom context~\citep{debnath2020designing}. Technological tools such as games, domain expert systems, and simulations offer an alternative approach to teaching requirements elicitation interviews by incorporating a digital interview partner. These tools can provide students with an interactive and engaging learning experience, allowing them to practice their skills in a controlled environment~\citep{daun2021survey}. Additionally, these tools can provide a standardized experience for all students, regardless of their partners' performance, ensuring that each student has an equal opportunity to develop their interviewing skills.

Emerging technologies have revolutionized the education ecosystem by creating new and innovative learning methods. For example, the rise of online learning platforms has enabled students to access educational content anywhere and anytime, making the learning experience more flexible~\citep{alraimi2015understanding}. 
Social robots have become a prevalent source of help in educational activities where success depends on regular practice and attentive supervision~\citep{belpaeme2018social}. Gamification techniques have been used to motivate and engage students by incorporating game design elements into education contexts~\citep{caponetto2014gamification}. Emerging technologies such as chatbots~\citep{paschoal2018chatterbot}, serious games~\citep{vega2009training,hainey2011evaluation,yasin2018design,garcia2020serious,ibrahim2019design,garcia2019experiences}, simulators~\citep{debnath2020designing} have also been used to support requirements engineering education. 

This paper explores the application of emerging technologies in requirements engineering education, aimed at resolving scalability concerns arising from the considerable human resources required for conducting elicitation interview practices. We propose \gensys{} as an extensible architecture for requirements elicitation interview training systems. 

Systems implementing \gensys{} feature two primary phases to help students practice and improve their elicitation interviewing skills. In the first phase, known as the interview session, the student takes the role of a requirements engineer while the system acts as a project stakeholder. Using a predetermined scenario, the system presents multiple choices for the following question for the student to ask at each turn of the interview and expects the student to choose the proper one. After the interview phase, the system passes to the tutoring phase, where it assesses the student's performance in the interview and provides feedback to improve their interview skills. The system presents the errors based on the choices of the student and allows the student to revisit the incorrect sections. It also offers contextual feedback to reinforce their learning. At the end of the feedback session, the system provides a behavioral analysis as potential soft skill improvement areas for the student. 

\gensys{} is a modular architecture that allows easy customization with alternative agent structures to accommodate the diverse needs and requirements of students and educators. We implement and evaluate two variants of \gensys{}: \sys{} and \tts{}. \sys{} employs an embodied robotic agent, enabling audio-visual interaction capabilities to better emulate a human-like interviewing context, albeit at a higher cost due to the required robotic hardware. In contrast, \tts{} adopts a virtual voice-based agent providing voice-only interaction, a cost-effective alternative offering a reduced human-like setup.

We conducted a user study with the students of a graduate level requirements engineering course to evaluate \sys{} and \tts{}. The participants rated both \sys{} and \tts{} favorably for perceived attitudes and usefulness without a significant difference. Regarding perceived ease of use, the participants' scores for \tts{} were higher than \sys{}, and this disparity is statistically significant. \tts{} is also found to be more engaging by the participants, although both of the systems rated with scores more than moderate levels. \sys{} yielded a notably higher learning gain compared to \tts{}, where the difference in learning gains between the two systems is significant. These results indicate that neither system dominates the other; rather, each system has its own distinct advantages and weaknesses.

The main contributions of this work are as follows:
\begin{itemize}
    \item We propose an extensible architecture for requirements elicitation training systems and demonstrate its applicability by implementing two instances of it with different agent structures and interaction modalities.
    \item We empirically evaluate the implemented interview training systems -- \sys{} with an embodied robotic agent and \tts{} with a virtual voice agent -- with the students of a graduate level requirements engineering course.
    \item We publicly share our system implementation and study materials in~\citet{binnur_gorer_2023_7861906}.
\end{itemize}

\gensys{} builds upon our previous work~\citep{gorer_roboreit}. Yet, it is designed to have a higher level of flexibility to support diverse agent configurations instead of a single agent configuration presented in our previous work, and it incorporates a behavioral feedback analysis component to support the soft skill improvement of students that is missing from our previous work. The user study to evaluate systems implementing \gensys{} is conducted with different subjects and research questions after their implementation. 

The rest of the paper is organized as follows. Section~\ref{sec:related_work} reviews the relevant related work for the use of emerging technologies in education in general and in requirements engineering (RE) education specifically. In Section~\ref{sec:system}, we outline the \gensys{} architecture together with the system descriptions of \sys{} and \tts{}. In Section~\ref{sec:evaluation}, we present our research questions, the design, and the findings of the user study. Section~\ref{sec:discussions} covers the discussion points, the limitations of the study, and the implications of our system on RE education. Section~\ref{sec:threats_to_validity} describes the threats to validity and how we address them. Finally, Section~\ref{sec:conclusion} provides the concluding remarks and future work.

\section{Related Work}
\label{sec:related_work}

Implementing innovative technologies in education encompasses diverse digital learning methods like e-learning, game-based learning, and artificial intelligence (AI) assisted learning~\citep{liaw2008investigating,sitzmann2011meta,ciolacu2018education}. Although these technologies have undergone continuous development over the last decade, the COVID-19 pandemic has expedited the adoption of technological tools in education. Remote learning has necessitated that these technologies should serve all students from various backgrounds and age groups worldwide. Consequently, having specialized solutions created with students' needs in mind has become imperative amidst this rapid digital educational transformation~\citep{zhao2021changes}. Moreover, the long-term applicability and accessibility of the systems turned out to be critical factors for institutions and instructors to adopt any technology in their education programs~\citep{pelletier20222022}. As a result, assessing a technological system's effectiveness in facilitating students' learning experience and outcome and its affordability by the institutions became essential to ensure long-term sustainability and effectiveness~\citep{lu2022usability}.

Requirements Engineering Education and Training (REET) plays a pivotal role in equipping students with the necessary skills to confront the industry's challenges. To provide students with practical exposure in real-world scenarios, it is imperative to merge academic knowledge with hands-on training~\citep{daun2021survey}. However, conventional pedagogical approaches often necessitate instructors' or peers' active involvement during practical training, potentially constraining the duration and repeatability of projects. Although several research has suggested using technical solutions to enhance the effectiveness of REET to address this challenge, they are relatively scarce given the advances in educational technology~\citep{daun2021survey}. We briefly outline the studies that propose using emerging technologies in REET. We only included the studies with an implemented solution and excluded proposal-only research that lack any quantitative or qualitative evaluation.

The majority of earlier research recommends software solutions to help REET learners standardize and validate the requirements stated during the elicitation processes. To assist in the creation of accurate and consistent software requirements specifications, \citet{garbers2006light} implement an interactive editing tool that enables the user to develop an IEEE-compliant requirements document. Using the tool, they target to teach how to improve requirements quality. A small number of students evaluated the tool and offered their feedback on its usability though no detailed analysis was provided on the efficacy of the tool. The issue of verifying the accuracy of the elicited requirements is addressed in~\citet{ogata2012training} and~\citet{kakeshita2015requirement}. \citet{ogata2012training} implement an automated tool that takes model-based requirements prepared using Unified Modelling Language (UML) and creates a prototyped product model. Their goal was to help instructors confirm the validity of students' requirements more easily. A group of students tested the tool and created more specific requirements. However, the missing details about the experiment, like the studied business case and the number of created requirements, limit the validity of the experiment results. A more sophisticated tool is offered in~\citet{kakeshita2015requirement} to help students determine whether their requirements analysis models are correct by comparing them to the one the instructor provided. This approach helps the students learn how to create more accurate models. The authors present a very rough analysis of survey results to measure the perceived usability and usefulness of the tool. \citet{liang2010experiences} remark on the need for remote collaboration of developers and customers while creating a requirements document. To teach this RE activity to the students, they develop a web-based wiki platform to encourage remote cooperation on the same requirements document. They present the feedback from students on the tool's usability but do not provide any evidence about its pedagogical efficacy. 

The use of digital games in REET suggests that games can be an effective and engaging method for teaching RE concepts and skills. These games are designed to simulate the challenges and complexities of RE tasks, such as eliciting and prioritizing requirements, and provide learners with opportunities to practice RE in a safe and supportive environment. A digital game is proposed in~\citet{vega2009training} to teach the stages of elicitation through the symbolic presentation of software requirements workshop activities. They aim to bring real-life elements of RE to the virtual world and offer an effective learning environment by playing. However, the study was limited in prototyping, and no user study was conducted. \citet{hainey2011evaluation} develop a game as a motivating and engaging platform to teach requirements collection and analysis procedures. As the pioneering study in REET to show a comparative evaluation of games to the paper-based traditional educational methodologies, they conducted a user study with a large and diverse group of students to assess the learning effectiveness of the game and students' perceptions of it. Likewise, an educational game is proposed in~\citet{yasin2018design} to teach the security requirements analysis process. Their user study primarily assesses the learning experience of the students in terms of the game's usability and utility as well as its motivational impact on the students. 

A three-dimensional (3D) serious game is created by \citeauthor{garcia2019experiences} to improve students' understanding and application of requirements engineering methodologies~\citep{garcia2019experiences,garcia2020serious}. The learning task in the study is to improve a set of pre-prepared requirements by inquiring more about them via elicitation interviews. The students were given the templates and asked to follow the game's guides to elicit better requirements. They conducted a large-group user study to measure the game's impact on student factors like motivation and satisfaction and the improvement of students' skills in preparing requirements with having a control group that did the same task without the game. Nonetheless, they did not present any statistical analysis of the game's effectiveness results. Another similar serious game was developed in~\citet{ibrahim2019design} to create an interesting and engaging learning platform, consequently aiming to improve students' understanding of the RE concepts. They aim to offer diverse experiences with multiple scenarios by having multiple levels within the game. However, only one level of the game is implemented, and pilot testing is applied with a few students to obtain feedback on the game's usability. The utilization of virtual reality technology for UML modeling in a 3D environment is demonstrated in~\citet{ochoa2019incorporating}. However, due to the lack of comprehensive user research, the suitability of the system for REET could not be assessed.

In a pioneering study outlined in~\citet{nakamura2014requirements}, the importance of incorporating a simulated stakeholder in requirements elicitation training is emphasized. The researchers designed a domain expert system that could monitor students' chat messages as they worked on eliciting requirements for a given project. The system intervened as necessary, encouraging students to ask questions and providing answers by conducting keyword-based searches on its extensive domain knowledge. Unfortunately, the effectiveness of the system was not adequately demonstrated in the user study because the scenarios used were too simplistic and did not require the students to seek assistance from the domain expert system. \citet{paschoal2018chatterbot} developed a prototype chatbot intending to aid students in improving their ability to elicit requirements. The chatbot was designed to adapt its responses to the student's expertise level, thereby appropriately customizing the complexity of the answer. Despite conducting an evaluation study to showcase the system's effectiveness compared to an out-of-context chatbot, the researchers utilized an unremarkable measurement that may threaten the study's validity. Another chatbot-based interview simulator is developed in~\citet{laiq2020chatbot}. The authors use cloud-based AI systems that can recognize the questions of the users provided in natural language and answer the questions based on the provided context. However, the authors did not include a comprehensive experimental study to provide more information about the simulator's operation and efficacy. \citet{debnath2020designing} propose an interview simulator with a multi-modal conversational agent to allow students to practice elicitation interviews. The simulator can also evaluate users' responses and provide a report at the end of the interview. While their approach is a significant contribution to the literature in addressing the characteristics of real requirements elicitation interviews, the prototype they have developed falls short of their original proposal in terms of interaction and dialogue generation. Their preliminary analysis shows that the participants who used the simulator before interviewing a human fictional customer made fewer mistakes than those who did not use the tool. In~\citet{konlog2023reit}, a web-based application is proposed to help with the creation of efficient training plans that are in accordance with the resources the instructor has. The application can accommodate different training programs, like our training schemes, and make them all available for the user to choose from in accordance with their needs.

The existing literature shows that although there have been certain advancements in the incorporation of technology within REET, such efforts remain relatively limited when contrasted with other educational domains, such as K-12 and higher education~\citep{leoste2021perceptions,timotheou2023impacts}, both in terms of the quantity of conducted studies and the diversity of explored technologies. Most studies in REET perform pilot user studies with students to assess suggested solutions, revealing insightful information about the students' preferences. Nonetheless, despite the importance of proving the efficacy of the suggested solutions through controlled experiments, many studies frequently lack this component, as noted in~\citet{daun2017common}. Furthermore, none of them explain the long-term applicability and utility of the suggested technology in REET-related education programs. In our study, we offer reproducible systems and evaluate them in terms of both user preferences and system effectiveness. Our purpose is to make it simple for instructors to assess the presented systems by taking into account their benefits and drawbacks in relation to the education objectives and available resources of the institute.

\section{Interview Training System}
\label{sec:system}

We propose a modular and extensible system architecture for Requirements Elicitation Interview Training, named \gensys{}. The architecture is designed to support requirements elicitation interview training activities which employ two main phases: interview practicing and interview performance evaluation. In the first phase, the agent acts as a stakeholder and allows the interviewer to practice an elicitation interview. Following a pre-determined scenario, the system presents multiple options for the interviewer’s next question at each interview stage. The scenario progresses based on the interviewer’s responses. For each interview turn, the agent tracks and analyzes the interviewer’s responses and facial expressions. In the interview performance evaluation phase, the agent acts as a tutor and revisits each erroneous turn, highlighting the incorrectly selected options and the reasoning behind them. The interviewer is given the opportunity to correct their answers.


\gensys{} is built using the Robot Operating System (ROS) framework~\citep{quigley2009ros}. This approach simplifies the creation of a modular architecture, allowing different modules to communicate synchronously through the publisher/subscriber messaging protocol. \gensys{} enables the use of physical or virtual agents. The system controller can incorporate various agent features and behaviors. In this study, we implement two versions of this architecture, each with different agents and associated features: \sys{} is built as a multimodal interactive robotic system, whereas \tts{} is implemented with a virtual voice-only agent\footnote{The code repository link for the systems' implementations is available in~\citet{binnur_gorer_2023_7861906}.}. 

In this section, we first introduce the components of \gensys{}. We then describe its customized versions designed with a robotic agent \sys{} and a voice agent \tts{}. Following, we outline the interaction flow of the interview training process and each step's functionality.

\subsection{System Architecture}
\label{sec:modarch}
\begin{figure}[htbp]
\centering
\includegraphics[width=0.98\textwidth]{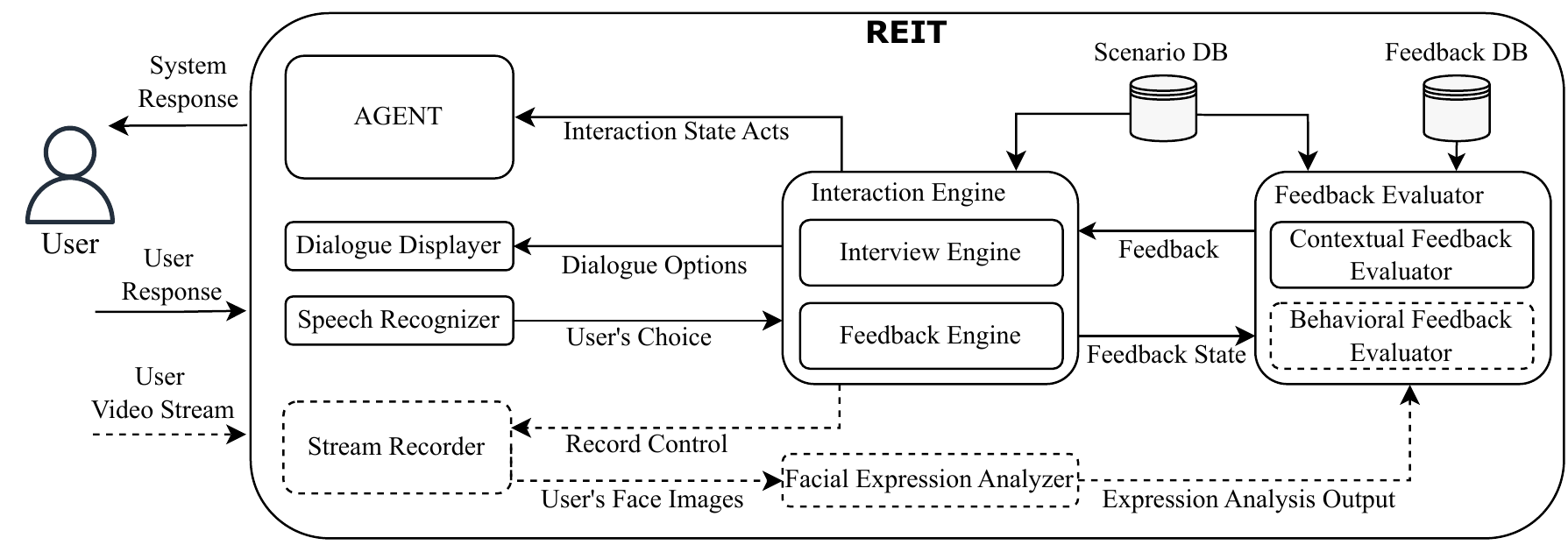}
\caption{The architecture \gensys{} for requirements elicitation interview training system. The dashed lines indicate optional inputs/outputs and modules.
}
\label{fig:sys_arch_reit}
\end{figure}

The modular interview training architecture \gensys{} is comprised of eight independent modules as shown in Figure~\ref{fig:sys_arch_reit}, namely \textit{Database, Speech Recognizer, Interaction Engine, Dialogue Displayer, Stream Recorder, Facial Expression Analyzer, Feedback Evaluator} and \textit{Trainer Agent}. The computation load is distributed across computing resources to enable real-time application and prevent any delays. We used two standard consumer laptops with modest memory and computing power (16 GB memory, 2.6 GHz 6-core CPU). The entire system is built to function autonomously, except for the speech recognizer module.

\paragraph{Database} includes scenario and feedback databases. The \textit{scenario database} is a collection of scenario files. An interview scenario is set up as a dialogue in which the requirements engineer and the stakeholder alternate responses. The requirements engineer has three options to choose from at each turn: two of them are manipulated with some mistakes that are commonly encountered in real elicitation interviews as described in~\citet{bano2019teaching}, and the other is the correct response. The selected option of the requirements engineer determines the stakeholder response and subsequent options for the requirements engineer for the following turn. As a result, the responses from the requirements engineer determine how the conversation flows across all possible dialogue paths in the scenario. The scenarios are created using the Twine tool\footnote{https://twinery.org/} and exported to HTML format for deployment. The \textit{feedback database} comprises feedback texts associated with each distinct mistake type. For example, the system provides the feedback ``Please do not use technical jargon the stakeholder may not be familiar with" if the selected option contains technical jargon. We devise multiple feedback texts for each mistake type by rephrasing the text to avoid repeating identical content in each occurrence of a certain mistake.

\paragraph{Speech recognizer} module converts speech inputs into text format. It captures user responses during both the interview and feedback sessions. In our study, this module was executed in a Wizard of Oz fashion, where a human operator was included in the process to convert the user's unstructured speech input (\textit{User Response}) into the expected system input (\textit{User's Choice}). The human operator handles participant speech in its original form and converts it into the intended system input only when it corresponds with one of the available dialogue options without adding their interpretations or judgments. 

\paragraph{Interaction engine} controls the interaction flow by iterating over the interaction steps. The overall interaction, described in Section~\ref{sec:intflow}, is modeled as a finite state machine. Each state has its defined actions and next-state transition logic. The engine decides the next state based on the current state and the relevant system inputs and variables defined in the state's transition logic. The interaction engine has two internal sub-engines devoted to managing the two main phases of the interview training:
\begin{itemize}
    \item \textit{Interview Engine} manages the interaction in the interview session. It queries the scenario with the given user's response (\textit{User's Choice}) to retrieve the stakeholder response text, which is then sent to the agent module to be synthesized and spoken to the user (via \textit{Interaction State Acts}). Meanwhile, the following question is collected from the scenario. As the agent finishes speaking the stakeholder response, the next question (\textit{Dialogue Options}) is passed to the dialogue displayer, where the user will be shown the possible dialogue options.
    \item \textit{Feedback Engine} controls the flow of the feedback session. The engine iterates over each incorrect interview turn. To remind the user of the interview stage, the preceding stakeholder text and dialogue options are sent to the dialogue displayer together with the previously chosen option. Simultaneously, the contextual feedback text (\textit{Feedback}) is received from the feedback evaluator module. It is then delivered to the agent controller to be uttered by the agent. After the user makes a second attempt, the feedback state (\textit{Feedback State}), which contains the user's updated response, is delivered to the feedback evaluator. The returned feedback (\textit{Feedback}), indicating whether the user's second attempt is correct or incorrect, is communicated to the agent module and the dialogue displayer in order to notify the user both verbally and visually.
\end{itemize} 

\paragraph{Dialogue displayer} is a graphical user interface to communicate text-based input of \gensys{} to the user, presented on the system's display. At each interview turn, it shows the dialogue options to the user to select one from (see Figure~\ref{fig:optScreen_a}). When the user makes a selection, the option is highlighted with a yellow background to let them know that the system understood their selection, as shown in Figure~\ref{fig:optScreen_b}. The tool is also used in the feedback session to display the revisited interview turns with incorrect responses. For each incorrect interview turn, the stakeholder's prior response and the corresponding options are displayed, with the incorrect option highlighted in red (see Figure~\ref{fig:secondchance_a}). The user's choice is highlighted in yellow following their second attempt to reassure the user that the system has recognized their input. The yellow background of the selected option will change to green if it is chosen correctly this time (see Figure~\ref{fig:secondchance_b}). Otherwise, it is turned red, and the correct option is shown with a green background to inform the user. At the end of the overall session, the feedback evaluator module evaluates the user's behavioral and contextual performance during the interview and provides a comprehensive analysis to the user. The analysis's findings are verbally communicated through supplementary visualizations. The dialogue displayer is utilized to present these visualizations to the user.

\paragraph{Stream Recorder} captures the video stream from the user (\textit{User Video Stream}), which is then processed by the facial expression analysis module. The recorded content includes both the user's speech and frontal face images. OBS Studio\footnote{https://github.com/obsproject/obs-studio}, an open-source recording software, is utilized for capturing the video content of the interview session. This tool effectively captures, encodes, and records video content and provides a Python API for controlling the program inside \gensys{}. During each interview turn, the recording starts as soon as the dialogue options are presented to the user and continues through the user's evaluation of the available option, the utterance of the selected option, and receiving the corresponding response of the agent. The recording then stops, saving the data captured during that turn for later analysis to provide feedback on the user's behavioral performance. The start and stop requests (\textit{Record Control}) are managed by the interaction engine.

\paragraph{Facial Expression Analyzer} evaluates the recorded face images of the user to determine their emotional state during each interview turn. This information is then used to provide feedback to the user, helping them to improve their soft skills. The goal of the emotional analysis is to give the user an understanding of their emotional state during the interview and highlight the significance of emotional control during requirements elicitation interviews. For the representation of the emotional states, we use Russell's circumplex model of emotions~\citep{russell1980circumplex}. This model describes emotions in terms of valence and arousal dimensional spaces. Valence refers to the degree of positivity or negativity associated with a particular emotion, while arousal refers to the level of excitement or calmness experienced. The FaceChannel library~\citep{barros2020facechannel} is utilized for the automated analysis of emotional states in arousal and valence dimensions. The library performs real-time analysis of facial expressions, making it an efficient and practical solution for our experimental setup with limited computational resources. It predicts the user's emotional state from the given image frames. The predicted values are represented on a continuous scale ranging from -1 to 1, with negative values indicating a low level of arousal or valence and positive values indicating a high level.

\paragraph{Feedback Evaluator} assesses the user's performance throughout the interview practicing. It has two components:
    \begin{itemize}
    \item \textit{Contextual Feedback Evaluator} evaluates the user's responses for each of the interview turns. If the user's choice is incorrect, that interview turn is saved with the preceding stakeholder text, user dialogue options, and the selected incorrect option. The mistake in the incorrect option and the corresponding feedback text for that mistake type are retrieved from the scenario and feedback databases, respectively. The evaluation results are communicated to the feedback engine in \textit{Feedback} message. During the feedback session, the feedback engine goes through each incorrect turn by displaying the interview turn and requests the agent module to mention the associated feedback verbally.
    
    The contextual feedback evaluator module is also utilized to evaluate the accuracy of the user's second attempt (sent by \textit{Feedback State}) provided upon feedback. If the second attempt is also inaccurate, no additional feedback is given now. The evaluation result, indicating whether the user's second attempt is correct or incorrect, is communicated to the feedback engine, which notifies the user both verbally and visually through the agent and the dialogue displayer.
    
    At the end of the feedback session, an overall analysis of the contextual performance is presented to the user. The accuracy of the user's choice per each interview turn is visualized and presented to the user as shown in Figure~\ref{fig:contextual_analysis_1}. Blue ticks indicate the right choice, while red exclamation marks indicate the wrong choice. The second attempt's results are also displayed to notify the user if an erroneous turn is fixed during the feedback session. Subsequently, the number of mistakes per category is shown for the interview and feedback session as in Figure~\ref{fig:contextual_analysis_2}. The incorrect choice in each turn is manipulated with mistakes belonging to one or two categories. Hence, the total number of mistakes per category is greater or equal to the number of incorrect turns. Our goal with this analysis is to help the user to identify the mistake categories with which they struggle and monitor their development upon obtaining feedback. The accompanying speech texts, along with the visualizations in Figure~\ref{fig:contextual_analysis}, are sent to the feedback engine to be transferred to the agent module and the dialogue displayer. 

    \begin{figure*}[ht!]
        \centering
        \begin{subfigure}[t]{1.0\textwidth}
            \centering
            \includegraphics[width=0.8\textwidth]{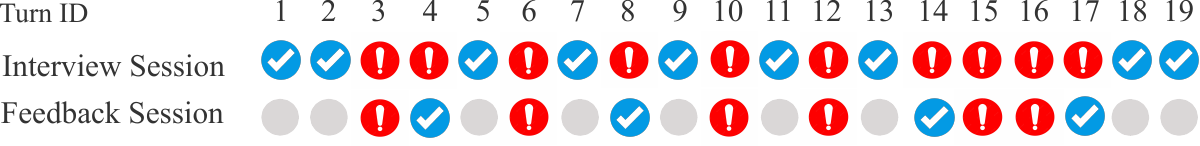}
            \caption{The state of user choice for each interview turn is displayed as correct or incorrect by blue ticks and red exclamation marks, respectively. The accuracy of user's second attempt in the feedback session is also displayed for the incorrect turns.}
            \label{fig:contextual_analysis_1}
        \end{subfigure}%
        
        \begin{subfigure}[t]{1.0\textwidth}
            \centering
            \includegraphics[width=0.6\textwidth]{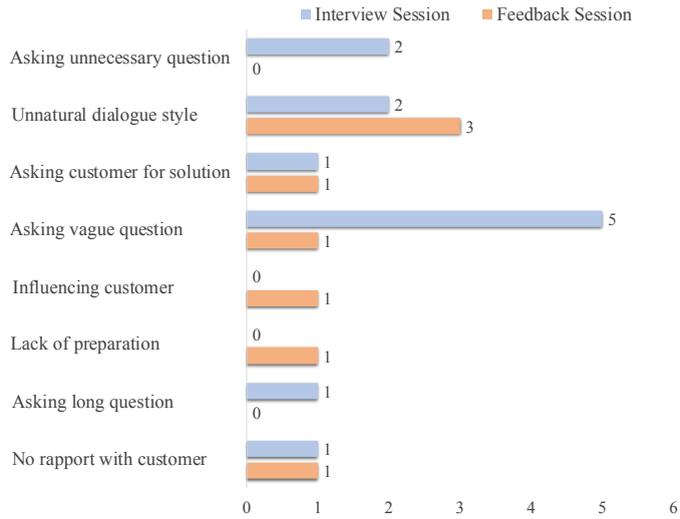}
            \caption{Number of user mistakes per mistake class in the interview session (in their first attempts) and feedback session (in their second attempts).}
            \label{fig:contextual_analysis_2}
        \end{subfigure}
    \caption{Visual presentation of the overall contextual performance analysis of a sample user at the end of the session.}
    \label{fig:contextual_analysis}
    \end{figure*}
    
    \item \textit{Behavioral Feedback Evaluator} takes the output from the facial expression analyzer and uses it to evaluate the user's behavioral performance. The module processes the results, which include the user's emotional state represented in terms of pleasure and activation, and computes the median values for each interview round. These values are depicted on the 2D circumplex model as in Figure~\ref{fig:circumplex_display}. The categorical emotions are placed on the circumplex to help the user better understand the representation of emotions in the dimensions of pleasure and activation. While conducting a requirements elicitation interview, it is important to avoid communicating negative emotions to the stakeholder. These emotions, such as nervousness, stress, upset, sadness, depression, and boredom, can negatively impact the interaction and decrease overall performance. Hence, the regions containing the negative emotions are indicated specifically on the circumplex. Emotional control is important for maintaining a professional and pleasant interaction during the interviews. It is desired for the analyst to display expressions with slightly positive emotions around neutral activation levels. To make it easier for the user to understand the agent's discussions around this topic, the desired emotional states are visually emphasized on the circumplex by highlighting the target region.
    
    In order to provide a more informative analysis across the turns, the descriptive statistics of pleasure and activation values are calculated and visualized as shown in Figure~\ref{fig:dimensional_analysis}. The median value provides a measure of central tendency, and the interquartile range ($25^{th}$ to $75^{th}$ percentile) shows the spread of the data. This helps the user to understand how their emotional expressions changed during the interview and identify any specific turns where the emotional expressions deviated from their typical pattern. This analysis could provide insights into the user's emotional control during the interview and helps them to identify possible shortcomings for improvement.

    \begin{figure*}[htbp]
        \centering
        \begin{subfigure}[t]{1.0\textwidth}
            \centering
            \includegraphics[width=0.4\textwidth]{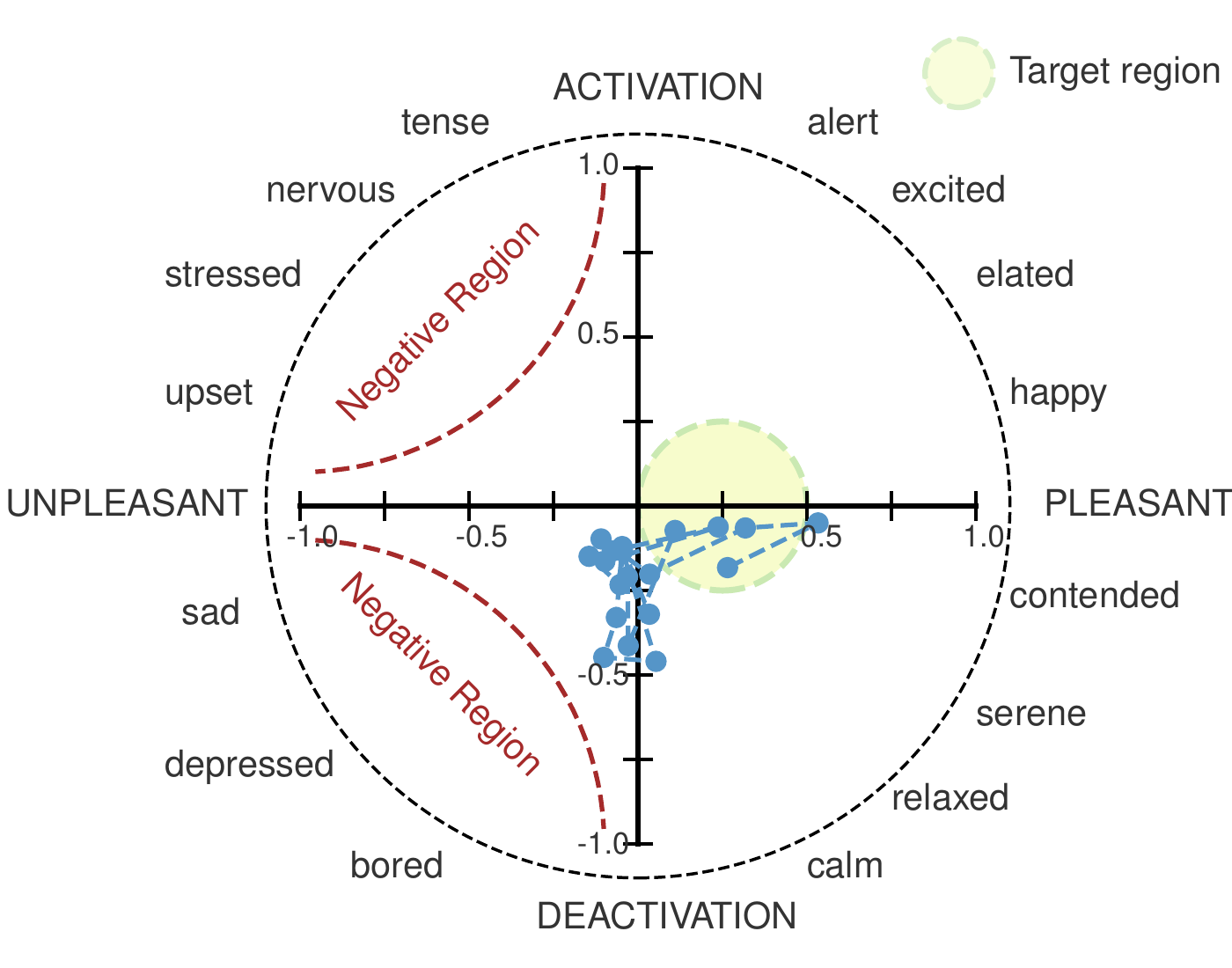}
            \caption{User's facial expressions during each interview turn are displayed on the 2D circumplex model with dimensions of pleasure and activation. The median value is used as the central value to represent the overall emotional state during each turn. The blue circles show the median values of emotional intensity per each interview round.}
            \label{fig:circumplex_display}
        \end{subfigure}%
        
        \begin{subfigure}[htbpΩ]{1.0\textwidth}
            \centering
            \includegraphics[width=0.5\textwidth]{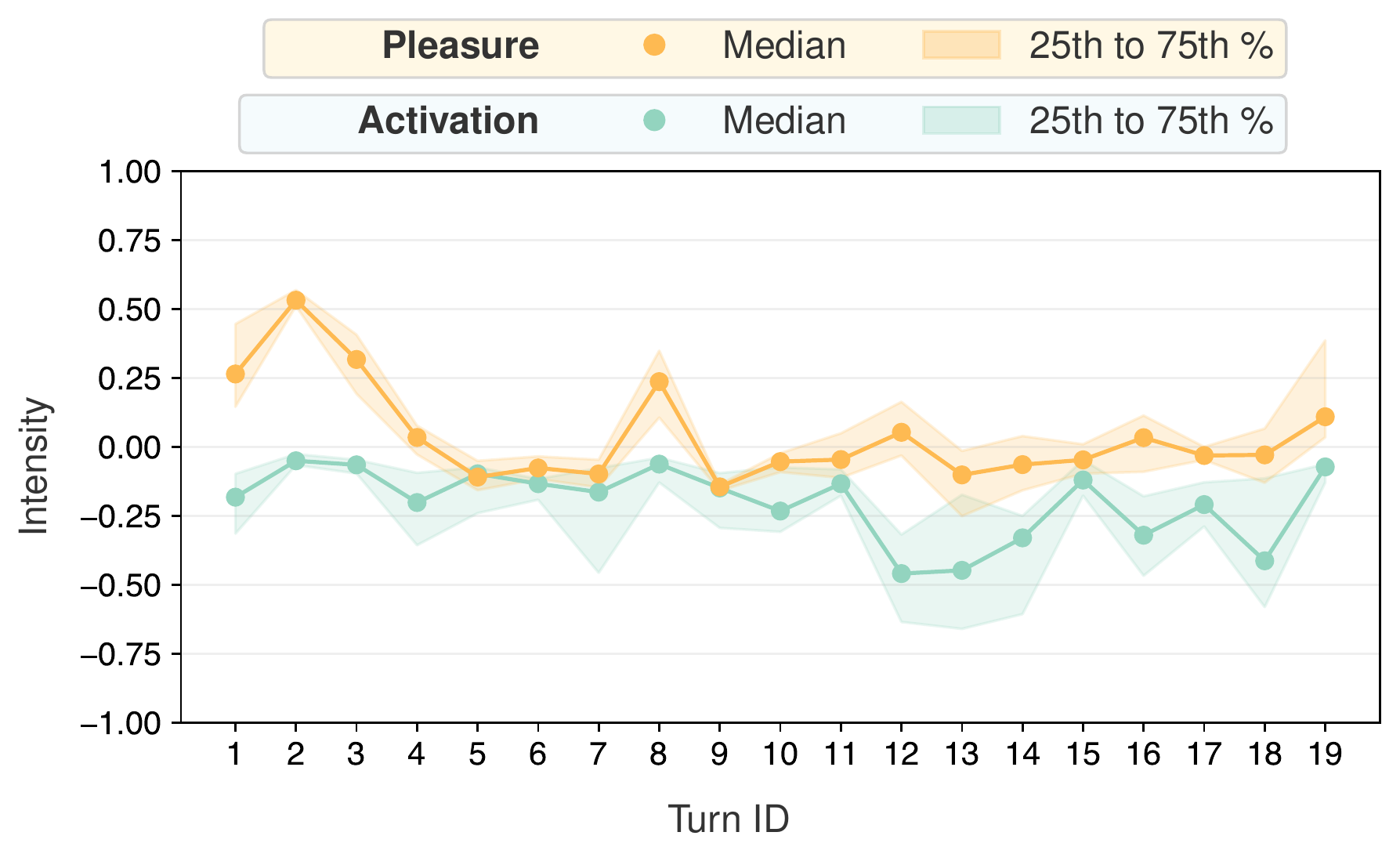}
            \caption{The visual representation of the descriptive statistics of the emotional expressions in the dimensions of pleasure and activation. The intensity value in each dimension ranges from -1 to 1. The median values and 25\% to 75\% percentile range are provided across all the interview turns.}
            \label{fig:dimensional_analysis}
        \end{subfigure}
    \caption{Visual presentation of the overall behavioral performance analysis of a sample user at the end of the session.}
    \label{fig:behavioral_analysis}
    \end{figure*}
    
\end{itemize}

\begin{figure}[htbp]
    \centering
    \includegraphics[width=0.5\textwidth]{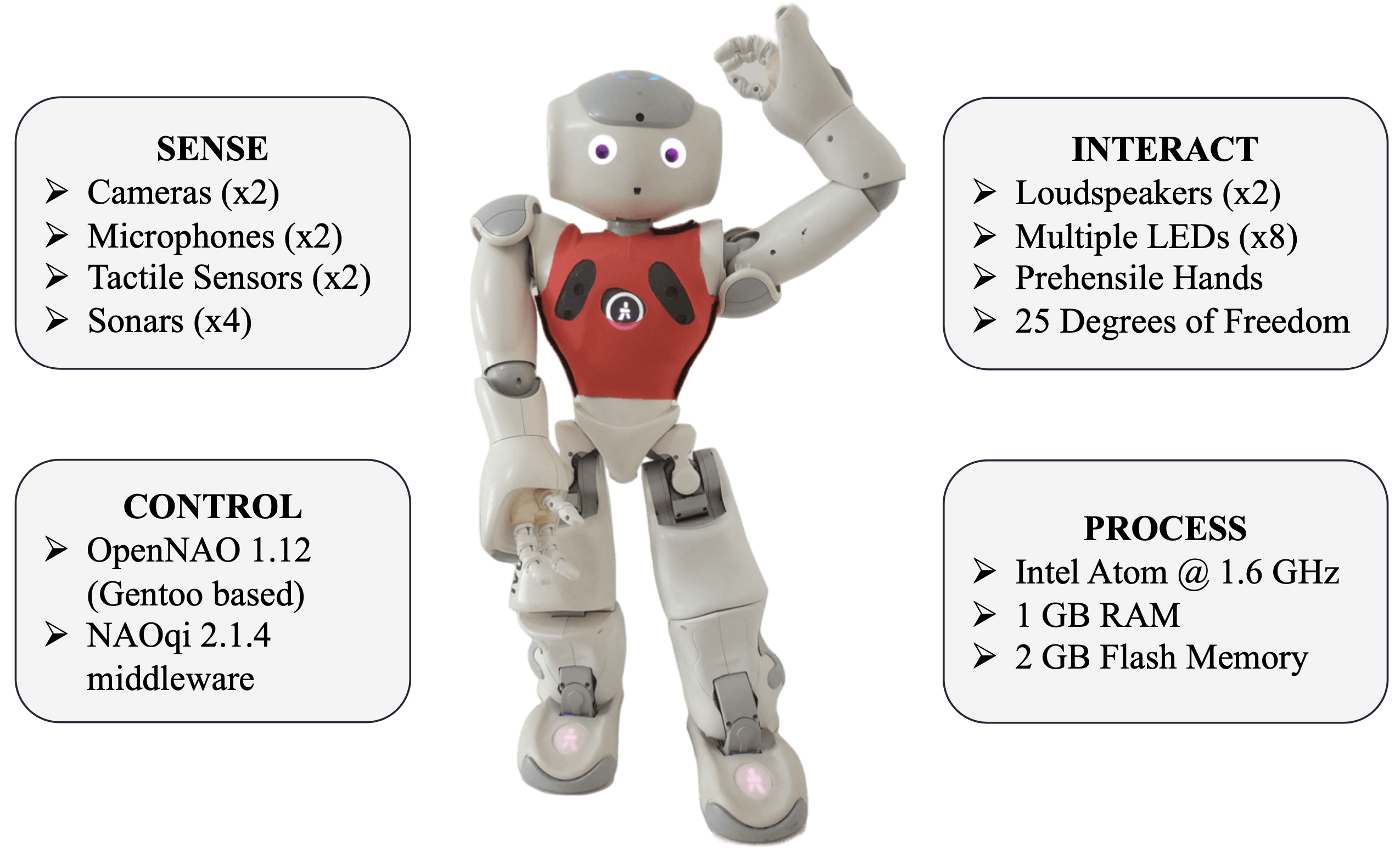}
    \caption{The Nao robot used in the robotic agent module.}
    \label{fig:nao}
\end{figure}

\begin{figure*}[ht!]
        \centering
        \begin{subfigure}[t]{0.45\textwidth}
            \centering
            \includegraphics[scale=0.8]{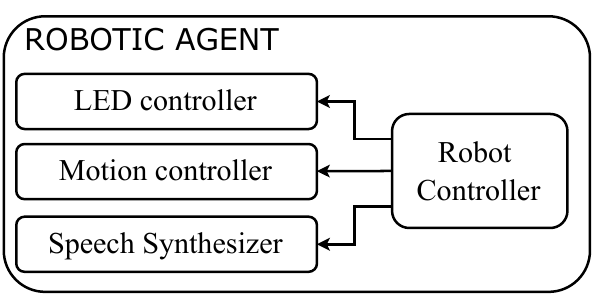}
            \caption{The robotic agent module.}
            \label{fig:sys_roit4re}
        \end{subfigure}
        ~
        \begin{subfigure}[t]{0.45\textwidth}
            \centering
            \includegraphics[scale=0.8]{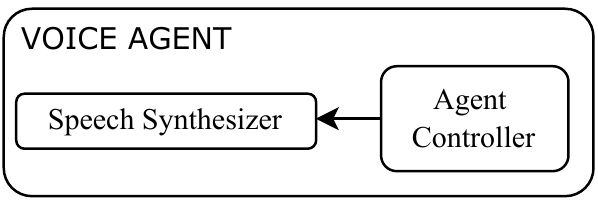}
            \caption{The voice-based agent module.}
            \label{fig:sys_voit4re}
        \end{subfigure}
    \caption{The agent module of \gensys{} is implemented with a robotic agent and a voice-based agent for \sys{} and \tts{}, respectively.}
    \label{fig:sub_systems}
\end{figure*}

\paragraph{Trainer Agent} module is a versatile and adaptable component designed to interact with users in various ways. Its primary objective lies in providing users with an intuitive interactive experience while efficiently communicating system actions. In this study, the module's implementation extends to both embodied and virtual forms, accommodating physical robotic agents as well as voice-based conversational agents. However, the module is designed to be flexible to support other sorts of agents. For instance, it can readily integrate with virtually embodied 3D characters within virtual reality platforms, permitting customization to cater to distinct user experiences. This module's extensible design further facilitates the incorporation of additional components, such as speech-driven gesture generation and user-adaptive emotional expressions, to enhance the agent's interactive capabilities. These additional features can be orchestrated by the interaction engine through state actions.
In our study, we implement two versions of the trainer agent module with different levels of interaction capabilities.
\begin{itemize}
\item \textit{Robotic Agent} is implemented with a physically embodied robotic platform, the Nao robot, which is shown in Figure~\ref{fig:nao}. As the most widely used humanoid bipedal robot for academic research, Nao is effective, affordable, and simple to program~\citep{gouaillier2008nao}. Its appealing design, expansive sensing, and acting capabilities make it sound for social robot-human interaction. It is 58 cm tall, mobile, and has auditory, visual, and tactile senses. Nao has two video cameras in the forehead and mouth, providing images with resolution up to 1280x960 at 30 frames per second. In our study, we used the $4^{th}$ version of Nao, which has the embedded Linux-based operating system OpenNAO and middleware NAOqi 2.1.4\footnote{http://doc.aldebaran.com/2-1/index.html}. Naoqi is the main program of Nao which provides a low-level programming interface to control the joints of the robot, adjust LEDs color and intensity, and manage the built-in text-to-speech component.

The robot controller module is depicted in Figure~\ref{fig:sys_roit4re}. It is responsible for controlling the actuators on the robotic platform in order to carry out the actions requested by the interview engine (\textit{Interaction State Acts}). The module manages the robot's speech synthesizer and the various sub-controllers of the robot's actuators, such as LEDs and motors. The robot controller communicates with the interaction engine to receive information about the current interaction state actions and converts them to the appropriate actions to be taken by the robot. The module is designed to provide seamless integration between the interview engine and the robotic platform, ensuring smooth and effective interactions between the user and the robot. The robot controller module is run on top of Naoqi middleware. Naoqi serves as a messaging interface between the Nao robot and the laptop and therefore is executed on both devices.

For the speech synthesizer module, we used the built-in functionality of the Nao robot. This module converts text into speech and plays it through the robot using the selected voice with customizable control commands such as pitch and speed. Besides speech, the robot has been programmed to use non-verbal cues such as head movements, gaze direction, and body gestures to convey social cues and enhance the overall interaction experience with the user. Emphatic gestures are created to express joy and despair in response to the users' correct and incorrect choices in their second attempts during the feedback session. Likewise, eye blinking is implemented using the eye LEDs of Nao to increase perceived liveness. These accompanying social skills aim to make the robot's communication more natural and engaging, leading to a better user experience and improved performance during the training.
\item \textit{Voice Agent} is designed to rely solely on speech-based communication without using any virtual or embodiment presence. It utilizes a speech synthesizer module to convert the text into speech (see Figure~\ref{fig:sys_voit4re}). gTTS\footnote{https://gtts.readthedocs.io/en/latest/} is used as the text-to-speech library. The agent controller manages this library by transferring the text message provided by the interaction engine (via \textit{Interaction State Acts}) and playing it back to the user through a speaker. The voice agent has no appearance or embodiment. A static image representing the agent is shown on the dialogue display of the user while the agent is speaking.
\end{itemize}

\subsection{\sys{} and \tts{}}
\gensys{} is built upon our previous work~\citep{gorer_roboreit}, targeting a more flexible architecture that can adapt to different agent configurations, feature sets, and scenarios. Our previous observations align with the common ``no size fits all" concept in the education literature, which recognizes that students may have different preferences and needs. Likewise, not all educational institutions may have the resources or infrastructure to support and apply a training system with an advanced technological component. 

The trainer agent module is designed to be flexible and adaptable to various types of implementations. This allows \gensys{} to be customized based on the specific requirements and use cases. 
The choice of the agent module may depend on the desired interaction modality and the available resources. By including or excluding different modules and choosing the appropriate agent module, \gensys{} can be tailored to meet specific needs and provide a customized experience for the user. We implement two different versions of \gensys{}; Robotic Requirements Elicitation Interview Trainer \sys{} and Voice-based Requirements Elicitation Interview Trainer \tts{}. Both implementations use the same core components, which are essential for a practical and effective training experience, such as the interaction engine and contextual feedback analyzer. The two systems fundamentally diverge in agent implementation and feedback evaluation aspects, which are influenced by the interaction mode the system offers.
\begin{itemize}
\item \sys{} provides an audio-visual interaction with the user. The embodied robotic agent used in this system can communicate with the user through speech and body gestures. Likewise, the user engages with the system via voice and video. The \sys{}'s audio-visual interaction enables the delivery of behavioral feedback. The modules required for the behavioral feedback analysis, including stream recorder, facial expression analyzer, and behavioral feedback evaluator, are activated in this system (see the optional components shown in Figure~\ref{fig:sys_arch_reit}). Despite these advantages, the system comes with a higher hardware cost, and the embodied robotic agent can interact with a single student at a time, making it less scalable then the implementation described below. 

\item \tts{} provides an interaction that is purely audio based. The agent lacks visual representation and has no visual modality in its communication. Video streaming is excluded in both ways of communication between the user and the system. Hence, the behavioral feedback analysis feature is not available in this version of \gensys{}. This design choice facilitates exploring and comprehending users' preferences on two distinct systems with different interaction modalities and feedback utility within the \gensys{} framework. \tts{} provides a more cost-effective and less resource-intensive solution compared to the robotic agent version. However, \sys{} provides more human-like communication and extended feedback evaluation with behavioral analysis.
\end{itemize}

\subsection{Interaction Flow}
\label{sec:intflow}
This section outlines the procedure for trainee interviewers utilizing the system. The session is divided into four distinct parts. To begin, the agent greets the interviewer and presents the system. The interview process then takes place, with the agent acting as a stakeholder for the project.
Following the interview, the feedback session starts, where the agent acts as a tutor and reinforces feedback on the problematic parts of the interview session. 
The system concludes the training session by providing an overall analysis, including a review of all interview turns and how much improved during the feedback session, and behavioral performance evaluation. Figure~\ref{fig:flow} illustrates the flow of interaction, with the interview and feedback processes clearly marked. The interaction steps are as follows:

\begin{figure}[htbp]
    \centering
    \includegraphics[width=0.7\textwidth]{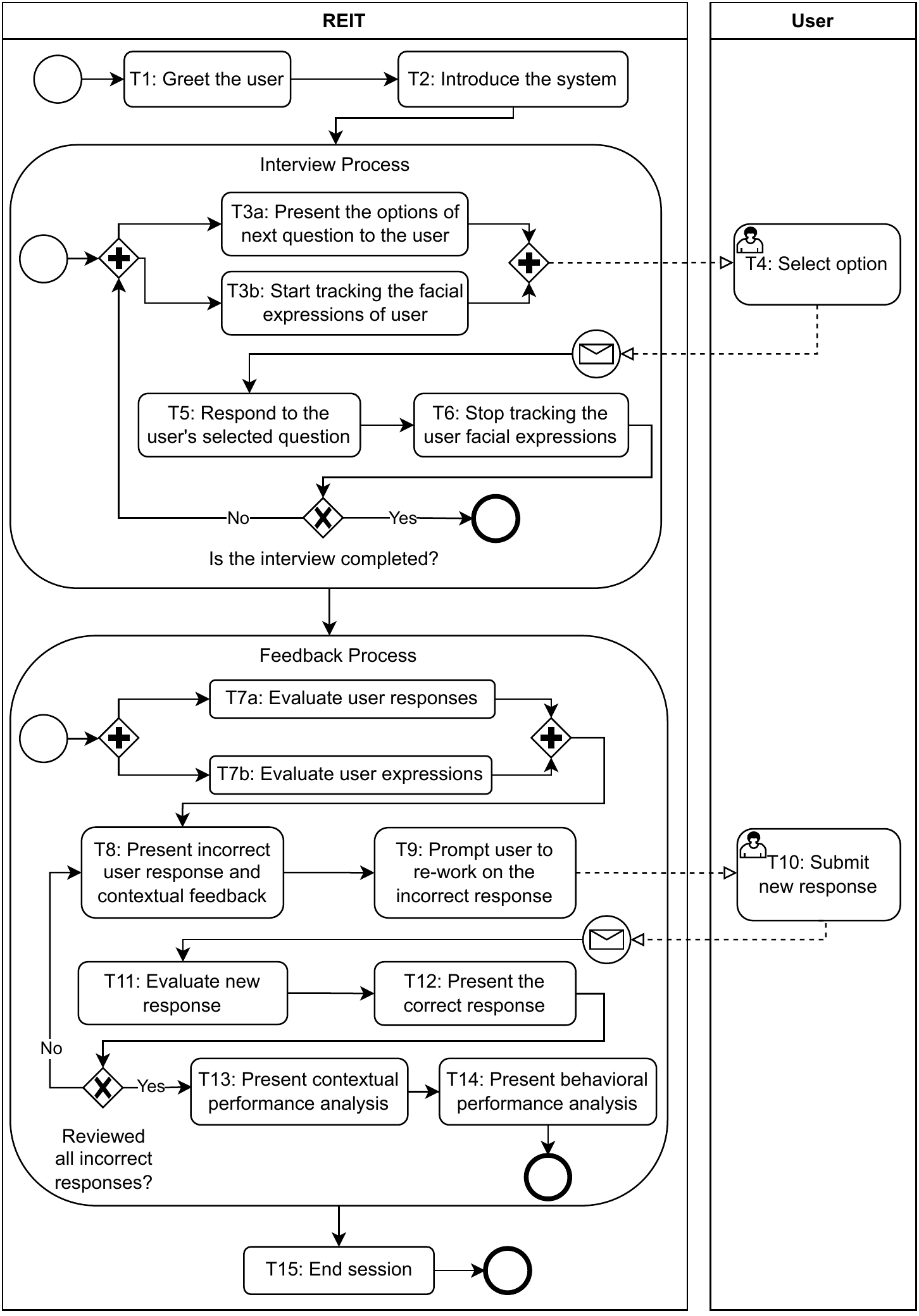}
    \caption{The interaction flow between the user and \gensys{}.}
    \label{fig:flow}
\end{figure}

\begin{description}
    \item[T1: Greet the user.] The agent greets the user and exhibits a praising affinity with the target of establishing a positive rapport with the user and creating a friendly and welcoming interaction environment to boost the user's acceptance of the system.
    \item[T2: Introduce the system.] The agent explains the function of the system and defines the user's role as the interviewer and its own role as the interviewee. The agent also provides a brief overview of the scenario for the session.
    \item[T3a: Present the options of the next question to the user.] At this stage, the system presents the user with a screen displaying potential questions for the interviewer to select from, as shown in Figure~\ref{fig:optScreen_a}. The user is prompted to select one of these questions to ask as the next question in the interview process. 
    \item[T3b: Start tracking the user facial expressions.] The agent begins to record the user's facial images. These images are later evaluated to provide feedback on the user's behavior during the interview.
    \begin{figure*}
    \centering
        \begin{subfigure}[t]{0.49\textwidth}
            \centering
            \includegraphics[height=1.25in, width=2.28in]{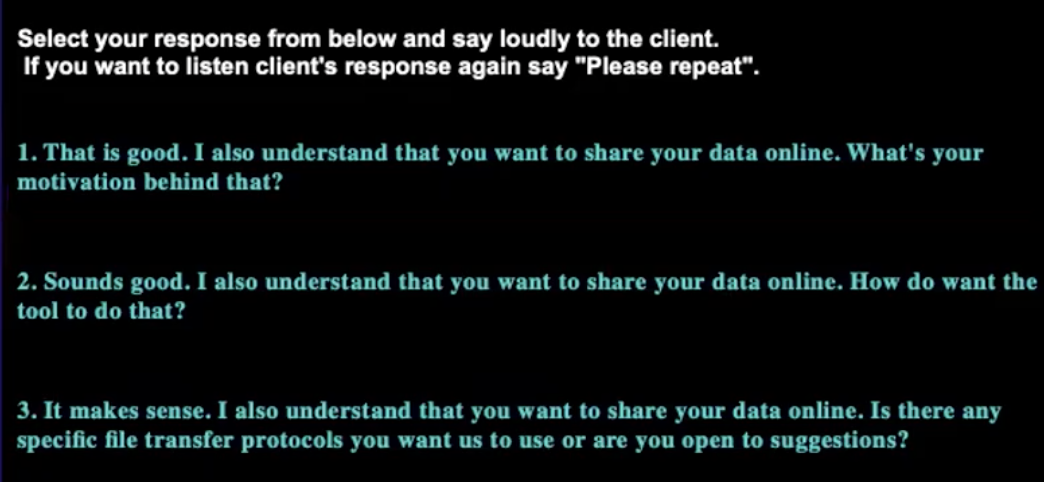}
            \caption{The agent presents the possible questions for the user to select from.}
            \label{fig:optScreen_a}
        \end{subfigure}%
        ~ 
        \begin{subfigure}[t]{0.49\textwidth}
            \centering
            \includegraphics[height=1.25in, width=2.28in]{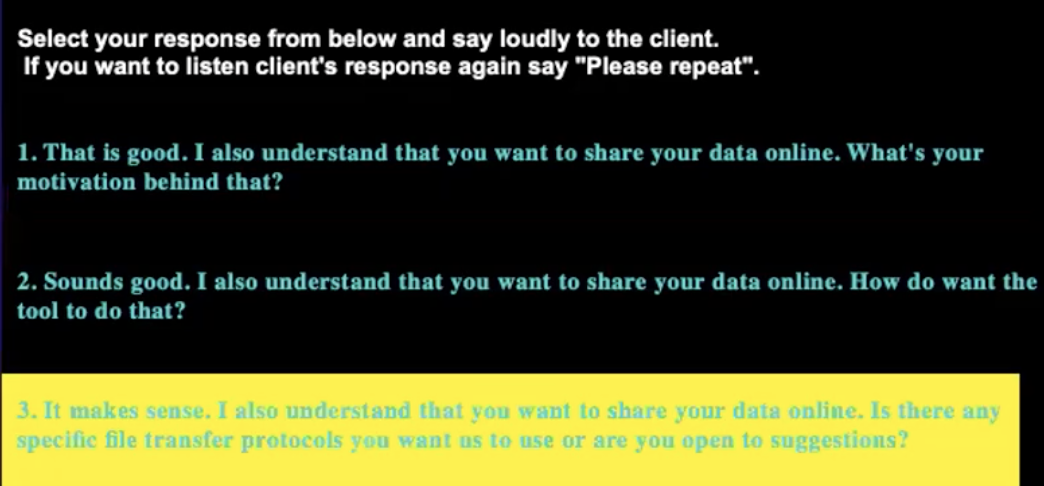}
            \caption{The agent confirms the user's choice by highlighting it.}
            \label{fig:optScreen_b}
        \end{subfigure}
    \caption{Samples of the interview training system's dialogue displayer states during the interview session.}
    \label{fig:optScreen}
    \end{figure*}
    \item[T4: Select option.] The user, who is acting as the interviewer, evaluates the presented options, taking into account the response of the stakeholder (the agent), the interview context, and the direction of the interview. The user then selects one of the options and speaks it out loud. The selected option is highlighted on the dialogue display to confirm the user's choice, as shown in Figure~\ref{fig:optScreen_b}.
    \item[T5: Respond to the user's selected question.] The system recognizes the input from the user and records the user's choice. The agent then responds to the question selected by the user according to the scenario flow. 
    \item[T6: Stop tracking the user facial expressions.] The agent stops recording the user's facial images for that turn. If there are more questions in the scenario, this loop continues with the next turn. Otherwise, the interview session is concluded, and the training continues with the feedback session.
     \begin{figure*}
    \centering
    \begin{subfigure}[t]{0.49\textwidth}
        \centering
        \includegraphics[height=1.35in, width=2.28in]{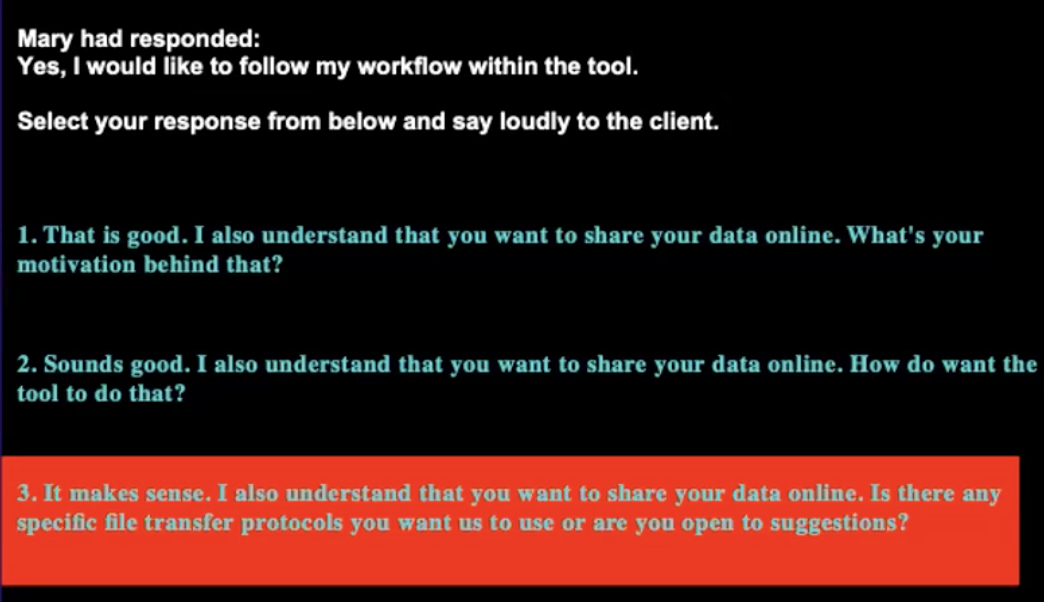}
        \caption{The agent highlights any incorrect responses given by the user and provides an opportunity for the user to correct them.}
        \label{fig:secondchance_a}
    \end{subfigure}%
    ~
    \begin{subfigure}[t]{0.49\textwidth}
        \centering
        \includegraphics[height=1.35in, width=2.28in]{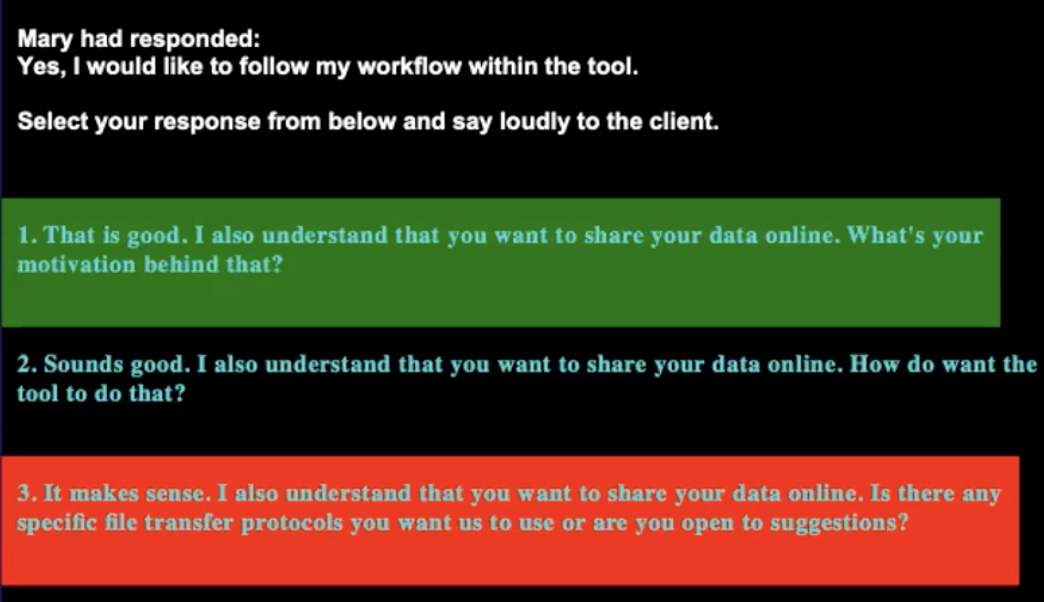}
        \caption{The agent notifies the user about the evaluation result of their second attempt on the incorrect response.}
        \label{fig:secondchance_b}
    \end{subfigure}
    \caption{Samples of the interview training system's dialogue displayer states during the feedback session.}
    \label{fig:secondchance}
    \end{figure*}
    
    \item[T7a: Evaluate user responses.] The system detects the turns with incorrect responses by comparing the user responses with the correct options in the specified scenario. The user is then notified by the agent that the feedback session has started.
    \item[T7b: Evaluate user facial expressions.] The system analyses the recorded facial images of the user to determine their level of arousal and pleasure during each interview turn. The system then computes the descriptive statistics and displays them, such as Figure~\ref{fig:behavioral_analysis}, to give the user an understanding of their emotional state during the interview. The analysis is presented to the user in the later stages of the feedback session.
    \item[T8: Present incorrect user response and contextual feedback.] The system revisits the incorrect turns that are evaluated in step T7a. For each turn, the agent's previous response to that turn and the available user options are presented. The incorrect response of the user is highlighted in red to remind their previous selection, as shown in Figure~\ref{fig:secondchance_a}. The agent provides the contextual feedback associated with the user's incorrect choice.
    \item[T9: Prompt user to re-work the incorrect response.] Upon providing the feedback, the system gives another chance to the user to identify the correct response.
    \item[T10: Submit new response.] The user re-evaluates the available options considering the interview context and the feedback provided. The user then makes a new choice and speaks it out loud.
    \item[T11: Evaluate new response.] The system compares the user's new choice with the correct option in the predefined scenario and verifies if the new response is correct.
    \item[T12: Present the correct response.] If the new response is correct, the system informs the user, as shown in Figure~\ref{fig:secondchance_b}. If the second trial of the user is also incorrect, the system displays the correct option with green while highlighting the others with red. For the second evaluation, no detailed feedback, like in step T8, is provided to keep the overall session length within limits.
    \item[T13: Present contextual performance analysis.] Upon reexamining all incorrect interview turns, the agent informs the user that the feedback session has ended. The agent provides the user's overall performance on each interview turn by visualizing the accuracy of user choice during interview and feedback sessions. Moreover, the system displays the number of incorrect answers and their mistake categories (as defined by \citet{bano2019teaching}) both for interview and feedback sessions. A sample visual presentation of the analysis is shown in Figure~\ref{fig:contextual_analysis}.
    \item[T14: Present behavioral performance analysis.] The system displays the behavioral analysis of the user's facial expressions for each interaction turns, as prepared in step T7b. In the meantime, the agent explains the goal and content of facial expression analysis and how this information might help the user better control their emotions during the interview.
    \item[T15: End Session.] After giving the user a sincere thank you for taking part in the training, the agent requests the user to complete the post-session questionnaires and concludes the session.
\end{description}

\section{Evaluation}\label{sec:evaluation}

This section presents the design, the execution procedure, and the empirical evaluation results of the user study. Our goal is to examine the comparative evaluations of the two systems: \sys{} with an audio-visual interaction interface involving an embodied physical robotic agent and \tts{} with an audio-only interaction having a virtual voice agent. We designed a user study wherein participants performed consecutive training sessions with both systems. During the sessions, the participants underwent an elicitation interview and received feedback on their performance, allowing them to learn from their mistakes. Half of the participants experimented with \sys{} first and then \tts{}, while the other half experimented with \tts{} first and then \sys{}. We collected users' perceptions of the systems' acceptability and engagement. We also measure the learning gain and interaction experience of the users. The following research questions (\textbf{RQs}) are addressed in our study:

\rqdef{RQ1: }\textit{How do \sys{} and \tts{} influence the learning gain of the participants?}\\
Both \sys{} and \tts{} systems are designed to deliver comparable interview and contextual feedback sessions, but they have different embodiment forms and interaction interfaces. Previous studies have suggested that robots may be more effective teachers than traditional training tools like web platforms or audiobooks, and that the physically embodied robots compared to their virtual equivalents can lead to measurable learning gains~\citep{han2005educational,leyzberg2012physical}. These studies, however, did not focus on adult learners in real-world educational setups like ours. This RQ investigates the impact of two systems on participants' learning gains in the context of requirements elicitation interview training. As the measure of learning gain, we use the normalized change in the number of mistakes made by participants between two subsequent interview sessions.

\rqdef{RQ2: }\textit{How do \sys{} and \tts{} influence the processing speed of the users during the interview session?}\\
The relationship between the diversity of interaction modalities and cognitive task load is an interesting and complex research topic. \citet{cao2009modality} suggest that exposure to diverse interaction modalities can lead to increased task load, as users need to process and adapt to different input and output channels. To investigate whether engagement in an audio-visual interaction with a physical robot might influence participants' task load differently from a voice-only interaction, we quantified participants' response speed during interview turns. In this context, we devise an in-context metric for processing speed, grounded in the principles of cognitive load theory in learning~\citep{kirschner2002cognitive}.

\rqdef{RQ3: }\textit{Does the participants' processing speed for a question vary based on their performance on the question?}\\
This RQ is focused on understanding whether there is a relationship between the participants' processing speed for a question and their performance on that question. Specifically, it aims to investigate whether participants' processing speed on the interview turns that they select the incorrect option differs from the turns they successfully respond to. The findings can provide insights into using our context-tailored processing speed measure to predict learners' performance during interviewing and improve the systems accordingly. 

\rqdef{RQ4: }\textit{How do \sys{} and \tts{} influence the perceived acceptance of the underlying system in the dimensions of \textbf{RQ4a:} perceived attitudes, \textbf{RQ4b:} perceived ease-of-use, \textbf{RQ4c:} perceived usefulness}?\\
An individual's intention to use a technological system is determined by their attitudes toward that technology and their perceived ease of use and usefulness. The technology acceptance model is a framework that helps to understand how and why individuals use technology. This RQ targets to evaluate users' attitudes and perception of ease-of-use and usefulness towards the underlying system used for requirements elicitation interview training. We employed an expanded version of the technology acceptance model questionnaire~\citep{yang2004s} to assess the participants' perceived acceptability of the two systems.
    
\rqdef{RQ5: }\textit{How do \sys{} and \tts{} influence the perceived and measured engagement levels during training with the system?} \\
Engagement is a crucial concept for the tutoring systems used in education because it can impact the effectiveness of the training process~\citep{trowler2010student,appleton2006measuring}. A tutoring system should keep the learner connected and engaged in order to maximize the learning outcome. If learners are not engaged in the educational session, they may not be paying enough attention, which can hinder their understanding of the subject. Contrarily, if learners are engaged in the tutoring session, they are more likely to actively participate and pay attention, which can enhance their learning experience and improve their understanding of the learning material. Engagement can also increase the learner's motivation and interest in the subject, leading to greater enjoyment and success in their studies. Hence, we would like to analyze how the proposed systems can succeed in engaging the users during the interview training sessions. To answer this RQ, we examine both perceived and measured engagement. To assess perceived engagement, we gather user responses on a questionnaire about how engaged they feel when using the system. For measured engagement, we extract the arousal levels from the users' speech samples that are collected during the interview sessions.

\rqdef{RQ6: }\textit{What are the relationships between individual user characteristics (i.e., age, gender, interview experience level, interview anxiety level) and perceived acceptance and engagement of the systems?}\\
This RQ concerns the relationship between individual aspects of the participants and their perceived acceptance and engagement of the systems. We aim to understand whether certain personal aspects correlate with perceived acceptance and engagement scores. We explored age, gender, interview experience level, and interview anxiety level as personal aspects.

\subsection{Study Design}\label{sec:study_design}
\begin{figure*}[htbp]
   \centering
	\includegraphics[scale=0.33]{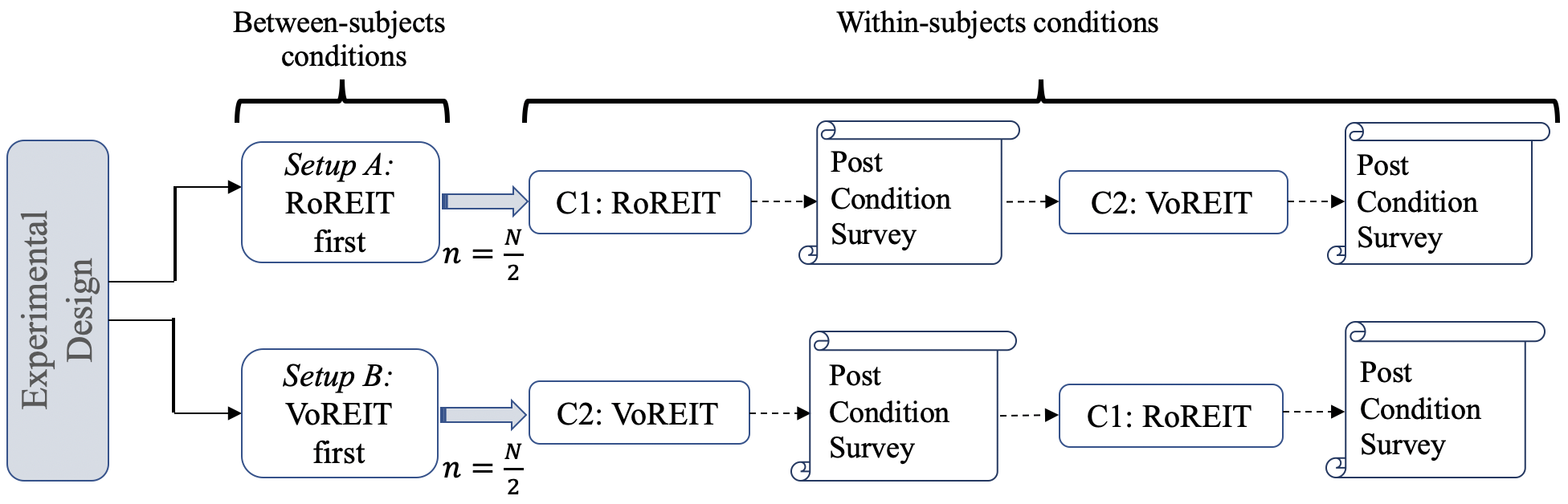}
	\caption{The experimental design of the user study.}
	\label{fig:study_design}
\end{figure*}

To investigate the user experiences on \sys{} and \tts{}, we create a user study employing a combination of between-subject and within-subject design by following the guidelines in \citet{hoffman2020primer}. In successive experiment sessions, participants use both \sys{} and \tts{}, so the system itself is the condition varying within subjects. As the between-subject condition, we created two configurations: \textit{setup A} and \textit{setup B}, where participants are randomly assigned to one of the two setups. In setup A, participants experiment with \sys{} first, followed by \tts{}, whereas in setup B, participants experiment with \tts{}, followed by \sys{}. Every experiment session includes training for requirements elicitation interviews with the given system, followed by completing the post-condition questionnaire. Different scenarios are used in each participant's successive sessions, and scenarios are allocated in random order across the participants. In this way, we ensure that no participant is exposed to the same scenario twice and that the system and scenario pairings happen in equal numbers. The overall study design is shown in Figure~\ref{fig:study_design}.

\subsubsection{Procedure}
\label{sec:procedure}

Due to the shift towards remote education right after the Covid-19 pandemic restrictions, the experiment is planned to be carried out online using a video conference tool. The experiment announcement, which includes information about the overall study, the informed consent form, technical requirements, and expected duration, is prepared in written and video formats and is shared on the social media channel for the requirements engineering class. Potential participants are encouraged to review the announcement and assess their availability first before expressing their interest in participating in the study. If a candidate meets the eligibility criteria and agrees to the informed consent form, which covers data protection and permission for video recording during the study, the experimenter reaches out to them to schedule an experiment time slot. Since English is the official language of education and is used by most business professionals in their work activities, including elicitation interviews, the experiment is prepared in English. Similarly, all materials, such as the study introduction, informed consent form, and administered questionnaires, are provided in English.
 
The target participant population's native language is not English, so they may not speak English fluently. Automated speech recognition systems (ASRs) demonstrate relatively reduced reliability when applied to non-native speakers in comparison to their performance with native speakers, particularly concerning unstructured speech~\citep{radzikowski2019dual,engwall2022wizard}. The structured format of the interview scenario adopted in our study could potentially rule out this issue by engaging ASRs with expected input sets. However, to expedite the prototyping of the systems, we opted for a Wizard of Oz approach. An experimenter assumes the role of the speech-to-text component, transcribing participants' speech into the expected system input.
The participants, however, are not aware of the experimenter's presence and are explicitly informed before the study that the experimenter will leave the environment once the study begins and that no help or assistance will be given during the session. This is to prevent any potential participant bias caused by participants' desire to look their best in front of the experimenter, which could cause them to not be sincere and genuine in their behaviors during the experiment. 

\begin{figure*}
    \centering
        \begin{subfigure}[t]{0.99\textwidth}
            \centering
            \includegraphics[scale=0.5]{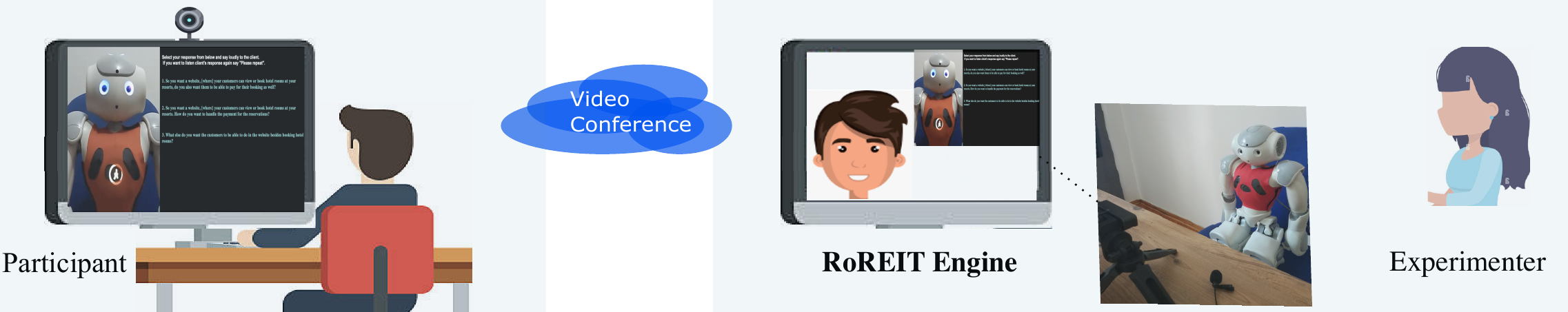}
            \caption{Experimental setup for \sys{}.}
            \label{fig:setup_roreit}
        \end{subfigure}
        ~
        \begin{subfigure}[t]{0.99\textwidth}
            \centering
            \includegraphics[scale=0.5]{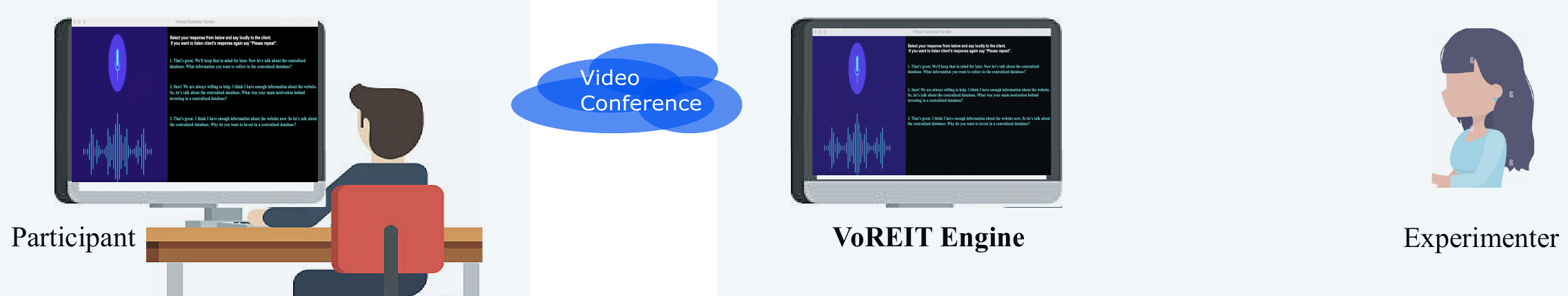}
            \caption{Experimental setup for \tts{}.}
            \label{fig:setup_voreit}
        \end{subfigure}
    \caption{Experimental setup for the user study.}
    \label{fig:setup}
\end{figure*}

The below steps are followed during the experimental process:
\begin{enumerate}
    \item[1)] Introduction: The experimenter dials into the video conference at the appointed time. When the participant joins, the experimenter describes the experiment's flow and responds to any questions they may have. Before the experiment begins, the participant is given the pre-experiment questionnaire to complete. The participant is asked to inform the experimenter once they are done with the questionnaire and ready to begin the study.
    \item[2)] Training: The participant performs the two consecutive sessions using \sys{} and \tts{} in the order of the assigned setup. The flow of a session is detailed below:
    \begin{itemize}
        \item[2a)]Readiness check: The participant is informed of the physical requirements of the experiment before the study begins.
        \begin{itemize}
            \item \sys{} condition: The experimenter makes sure the participant's face is visible and their voice can be clearly heard. The researcher then displays the dialogue displayer of \sys{}, which contains the live video of the robot and the dialogue options, on the shared screen (see Figure~\ref{fig:setup_roreit}). After getting the approval of the participant to start, the experimenter initiates \sys{} and notifies the participant that she will leave the environment. 
            \item \tts{} condition: There is no video interaction in this condition, and the participant conducts an audio-only interview with \tts{}. The experimenter ensures that the participant's camera is off and the participant's voice is audible clearly. The dialogue displayer of \tts{}, which includes the representative image of the text-to-speech agent and the dialogue options (see Figure~\ref{fig:setup_voreit}), is then presented on the shared screen by the experimenter. After receiving the participant's approval to start, the experimenter initiates \tts{} and informs the participant that she will leave the environment. 
        \end{itemize}
        \item[2b)]Session execution: Even though the participant is told to be left alone during the session, the experimenter stays in the environment to manage the system's speech-to-text component but remains completely invisible to the participant. She provides the selected option to the system after each verbal response of the participant. Through the process outlined in Section~\ref{fig:flow}, the participant conducts the session, which comprises the interview and feedback portions.
        \item[2c)]Post-condition questionnaire: The experimenter returns to the call again once the session is over and asks the participant to complete the post-condition questionnaire. 
    \end{itemize}
    \item[3)]Closure: The experiment is concluded once the two consecutive sessions and administered questionnaires are completed. The experimenter thanks the participant for joining and responds to any additional comments or questions from the participant.
\end{enumerate}

To assess the feasibility of the experimental procedure and the anticipated length of the overall experiment, we conducted two pilot trials with our colleagues. The duration of the experiment is estimated to be 60 minutes, divided as follows: 5 minutes for the introduction and completion of the pre-experiment questionnaire; 50 minutes for the two sessions, including completion of the post-condition questionnaire; and 5 minutes for the conclusion. We anticipate some variances in the overall experiment lengths, though, as the length of the feedback phase will depend on how many mistakes the participant makes.

The topological layout of the experimental setup of \sys{} is shown in Figure~\ref{fig:setup_roreit}. The study is placed in a room with a camera, a lapel microphone, the Nao robot sitting on a chair, and the laptop that runs the system components. For \tts{} setup, only the laptop to execute \tts{} system components including the virtual voice agent, is required, as illustrated in Figure~\ref{fig:setup_voreit}. For both systems, the experimenter is in the experiment room to observe the study for speech-to-text wizarding, but she is not visible to the participant. %
To reduce external distractions, participants are asked to take part in the experiment from a computer with a stable network connection in a well-lit, quiet, and comfortable setting. They are also required to have their camera turned on in \sys{} setup to enable audio-visual interaction.

\subsubsection{Materials}
\label{sec:materials}
The materials used in the experiment are available in our shared repository~\citep{binnur_gorer_2023_7861906}. 
\sectopic{Scenario.} We used two different scenarios in the experiments to minimize the learning effect over the course of the subsequent sessions. The scenario-system pairing is carried out in a randomized manner to alleviate any potential impacts of the scenario on the system. %
As one of the scenarios, we used the Cool Ski Resort scenario provided in \citet{ferrari2020sapeer}. A real-world use case, designing a website and its associated technological system for a resort company, is employed as the foundational scenario context. It is intended to elicit requirements for the website's design, social media integration, and database usage. We created another scenario, Cool Research Institute, by utilizing a comparable layout of the Cool Ski Resort scenario. In the second scenario, the institution's researcher, who is the stakeholder, wants to improve the processes for their research project. The goal is to create a new publication management system while also improving the existing data processing tool. The scenario aims to gather requirements for both the new publication management system and the enhancement of the existing data processing tool. This will include the addition of new features such as better data visualization and the ability to share data online. We carefully adjusted the dialogue portions for requirements engineer response options to keep the induced mistake counts and types closely similar between the two scenarios. Table~\ref{tab:scenario_mistakes} shows the total number of induced mistakes for each mistake class as described in~\citet{bano2019teaching}, which are counted over the requirements engineer's all possible responses in the given scenario. For both of the scenarios, the number of interaction turns for each interview party is set to a range of 15 to 19.

\begin{table*}
\centering
\footnotesize
\caption{The occurrences of mistakes induced in \textit{Cool Ski Resort} and \textit{Cool Research Institute} scenarios along with their associated mistake class.}
\label{tab:scenario_mistakes}
\begin{tabular}{lllrr}
\toprule
\multicolumn{1}{c}{\textbf{ID}} & \textbf{Mistake Class}                         & \textbf{Mistake Type}                                                      & \multicolumn{2}{c}{\textbf{Number of Occurances}}                                                                                                                                                      \\
                                &                                                &                                                                            & \multicolumn{1}{c}{\textit{Cool Ski Resort}} & \multicolumn{1}{c}{\textit{Cool Research Institute}} \\ \midrule
1                               & \multirow{2}{*}{\textit{Planning}}             & Lack of preparation                                                        & 4                                                                                       & 6                                                                                                 \\
2                               &                                                & Lack of planning                                                           & 3                                                                                       & 3                                                                                                 \\ \hline
3                               & \multirow{2}{*}{\textit{Question Omission}}    & Not identifying stakeholders                                               & 1                                                                                       & 1                                                                                                 \\
4                               &                                                & \begin{tabular}[c]{@{}l@{}}Not asking about existing\\ system\end{tabular} & 6                                                                                       & 6                                                                                                 \\ \hline
5                               & \multirow{5}{*}{\textit{Question Formulation}} & Asking long question                                                       & 3                                                                                       & 2                                                                                                 \\
6                               &                                                & Asking unnecessary question                                                & 7                                                                                       & 7                                                                                                 \\
7                               &                                                & Asking customer for solution                                               & 15                                                                                      & 15                                                                                                \\
8                               &                                                & Asking vague question                                                      & 32                                                                                      & 33                                                                                                \\
9                               &                                                & Asking technical question                                                  & 5                                                                                       & 3                                                                                                 \\ \hline
10                              & \textit{Order of interview}                    & Incorrect ending                                                           & 6                                                                                       & 6                                                                                                 \\ \hline
11                              & \multirow{2}{*}{\textit{Customer interaction}} & Influencing customer                                                       & 9                                                                                       & 10                                                                                                \\
12                              &                                                & No rapport with customer                                                   & 16                                                                                      & 18                                                                                                \\ \hline
13                              & \textit{Communication skills}                  & Unnatural dialogue style                                                   & 11                                                                                      & 13                                                                                                \\ \bottomrule
\end{tabular}
\end{table*}

\sectopic{Pre-experiment questionnaire.} We collect the participants' demographics, their opinions about the robots, and their perceived interview anxiety.
\begin{itemize}
    \item \textit{Demographics:} In order to gain a better understanding of the participant population, we gathered information on their general demographics, which included age, gender, current occupation, and years of working experience. In addition to these basic demographic questions, we also prepared questions using 5-point Likert scale to get participants' proficiency and experience levels in conducting requirements elicitation interviews. We asked about their previous training and experience with eliciting software requirements, as well as the number of times they had practiced and conducted real interviews. Despite our target participant group being students of a graduate level requirements engineering course, we wanted to gather the self-evaluation of their proficiency in conducting requirements elicitation interviews. We asked them to rate their conceptual understanding of the fundamentals of requirements elicitation interviews, as well as their confidence level for practicing the interviews.

    \item \textit{Negative Attitudes towards Robots:} We measured the participant's attitudes towards robots. If participants have a bias towards robots, this could impact the experiment result validity. The ``Negative Attitude toward Robot Scale" (NARS) measures people's anxiety towards robots. It is designed to gauge people's attitudes and beliefs about robots, including their level of confidence and trust in robots, their understanding of the capabilities of robots, how at ease they are interacting with robots, and the extent to which they believe robots will have a positive or negative impact on society. We used the NARS version provided in~\citet{syrdal2009negative}.
    
    \item \textit{Interview Anxiety Questionnaire:} When people feel anxious during an interview, it can impact how they perceive and experience the interview process, including their interactions with the interviewer. \citet{mccarthy2004measuring} developed a multidimensional measure of interview anxiety that provides a predictive anxiety level of a person, especially for selection interviews (e.g., job interviews). It typically includes a series of statements that assess the individual's level of nervousness or fear associated with the interview process. These involve the topics such as how confident they feel about their qualifications, how comfortable they are with being evaluated by others, and how much they anticipate experiencing anxiety during an interview. The measure provides an assessment of five interview anxiety dimensions: Communication, Appearance, Social, Performance, and Behavioral. We created a subset of the measure by reducing it to the items that are more relevant to the requirements elicitation interviews. Additionally, we changed the wording slightly to make the items more suitable for requirements elicitation interviews rather than the selection interviews. The adapted version of the measure is presented in Table~\ref{tab:anxiety_ques}.
\end{itemize}

\newcommand{\PreserveBackslash}[1]{\let\temp=\\#1\let\\=\temp}
\newcolumntype{C}[1]{>{\PreserveBackslash\centering}p{#1}}
\newcolumntype{R}[1]{>{\PreserveBackslash\raggedleft}p{#1}}
\newcolumntype{L}[1]{>{\PreserveBackslash\raggedright}p{#1}}

\begin{table*}[]
\footnotesize
\caption{The interview anxiety questionnaire adapted from \citet{mccarthy2004measuring}. Each item is rated on a 5-point Likert scale (1 - Strongly disagree to 5 - Strongly agree).}
\label{tab:anxiety_ques}
\bgroup
\def\arraystretch{1.1}
\begin{tabular}
{p{0.1cm}L{1.8cm}L{13.3cm}}
\hline
\textbf{ID} & \textbf{Anxiety Scale} & \textbf{Item}                                                                                                                                          \\ \hline
\rowcolor[HTML]{EFEFEF} 
1           & Communication          & I feel that my verbal communication skills are strong.                                                                                                 \\ \hline
2           & Social                 & \begin{tabular}[c]{@{}l@{}}While taking an interview, I become concerned that the interviewer will perceive me as socially awkward.\end{tabular}     \\ \hline
\rowcolor[HTML]{EFEFEF} 
3           & Performance            & \begin{tabular}[c]{@{}l@{}}I am overwhelmed by thoughts of doing poorly when I am in an interview situation.\end{tabular}                            \\ \hline
4           & Appearance             & \begin{tabular}[c]{@{}l@{}}Before an interview, I am so nervous that I spend an excessive amount of time on my appearance.\end{tabular}              \\ \hline
\rowcolor[HTML]{EFEFEF} 
5           & Communication          & During the interviews, I often can not think of a thing to say.                                                                                        \\ \hline
6           & Performance            & \begin{tabular}[c]{@{}l@{}}During an interview, I worry about what will happen if the interviewer is not satisfied with my performance.\end{tabular} \\ \hline
\rowcolor[HTML]{EFEFEF} 
7           & Social                 & \begin{tabular}[c]{@{}l@{}}I get afraid about what kind of personal impression I am making on interviews.\end{tabular}                              \\ \hline
8           & Appearance             & \begin{tabular}[c]{@{}l@{}}I often feel uneasy about my appearance when I am being interviewed.\end{tabular}                                        \\ \hline
\rowcolor[HTML]{EFEFEF} 
9           & Social                 & \begin{tabular}[c]{@{}l@{}}I become very uptight about having to interact socially with an interviewer.\end{tabular}                                 \\ \hline
10          & Social                 & I worry about whether the interviewer will like me as a person.                                                                                        \\ \hline
\rowcolor[HTML]{EFEFEF} 
11          & Performance            & \begin{tabular}[c]{@{}l@{}}In interviews, I get nervous about whether my performance is good enough.\end{tabular}                                    \\ \hline
12          & Communication          & \begin{tabular}[c]{@{}l@{}}I get anxious in the interviews that I am unable to express my thoughts clearly.\end{tabular}                             \\ \hline
\rowcolor[HTML]{EFEFEF} 
13          & Social                 & \begin{tabular}[c]{@{}l@{}}During an interview, I worry that my actions will not be considered socially appropriate.\end{tabular}                   \\ \hline
14          & Performance            & \begin{tabular}[c]{@{}l@{}}During an interview, I am so troubled by thoughts of failing that my performance is reduced.\end{tabular}                \\ \hline
\rowcolor[HTML]{EFEFEF} 
15          & Communication          & \begin{tabular}[c]{@{}l@{}}During the interviews, I find it hard to understand what the interviewer is asking me.\end{tabular}                       \\ \hline
\end{tabular}
\egroup
\end{table*}

\sectopic{Post-condition questionnaire.} Through a series of carefully designed and adapted questionnaires, we solicit feedback from participants on their interactions and experiences with the system in order to answer the research questions of the study.
\begin{itemize}
\setlength\itemsep{1em}
    \item \textit{Questionnaire on System Design:} We gathered information on the participants' perceived engagement level for the interview system. %
    The questionnaire is designed using a 5-point Likert scale.
    \item \textit{Technology Acceptance Model:} The technology acceptance model (TAM) was presented by \citet{davis1989perceived} to reveal two key attitudes that drive the adoption of information technology to accomplish a task: perceived usefulness and ease of use. According to TAM, an individual's attitude towards technology is influenced by their beliefs about its usefulness and ease of use. %
    TAM can help to understand user opinions and predict their intention to use it. This information can be useful for technology developers, as it can help them to improve the technology and make it more appealing to users. \citet{yang2004s} refined the original TAM by also taking into account the affective and cognitive aspects of attitude and provided an expanded version. In our research, we used their version, which includes the following sections:
        \begin{itemize}
        \setlength\itemsep{1em}
            \item[] a) \emph{Attitudes towards using} a technology can influence an individual's intention to use it.
            If an individual has a strong positive attitude towards technology, they will be more likely to use it, even if they perceive it as difficult to use or useless at all. On the other hand, if they have a weak positive attitude towards the technology, they may be more hesitant to use it, even if they perceive it as easy to use or useful.
            \item[] b) \emph{Perceived ease-of-use} refers to an individual's perception of how easy it is to use the technology. It is one of the key elements affecting a person's decision to use technology. The evaluation of ease-of-use may consider both physical and mental effort and an effort to learn how to use the technology.
            \item[] c) \emph{Perceived usefulness} is defined as the degree to which a person believes that using a particular system would enhance their job performance. It relates to how much the system can improve a person's task efficacy and how valuable the system is in connection to the content or goal of the task.
        \end{itemize}
    \item \textit{General remarks:} We asked participants about what they liked and disliked about the system and their suggestions for further improvements as open-ended text. We discuss their responses in Section~\ref{sec:discussions}.
\end{itemize}

\subsubsection{Dependent Variables}
\label{sec:dependent_vars}
The dependent variables arising from the RQs are 
\textit{learning gain} (RQ1), \textit{processing speed} (RQ2), \textit{turn-specific processing speed} (RQ3) \textit{perceived acceptance} (RQ4 and RQ6), \textit{perceived engagement} (RQ5 and RQ6), and \textit{measured engagement} (RQ5). Their formal definitions are given below.

\paragraph{Learning gain} is a measure of the improvement in learning that occurs as a result of a particular instruction or intervention. It can be used to assess the effectiveness of educational programs and materials, and has been applied in a variety of settings, including K-12 education, higher education, and workplace training~\citep{merchant2014effectiveness,roohr2017investigating}. Learning gains have been quantified in a variety of methods. Most frequently, this has been accomplished by tracking changes in the test results of the students during the course of instruction or training sessions. \citet{hake1998interactive} introduced normalized gain to advocate a consistent analysis over diverse student populations with widely varying initial knowledge states. It is calculated by dividing the amount students learned by the maximum amount they could have learned, as shown below:
\begin{equation}
\ev{g} = \dfrac{\ev{post}-\ev{pre}}{100-\ev{pre}}
\label{eq:learning_gain}
\end{equation}     

We adapted this measure to our study by considering the number of errors made by participants in a session as the success indicator. The difference between the number of mistakes in two consecutive sessions is used as how much each participant learned in the first session. In Equation~\ref{eq:learning_gain}, 100 is specified as the optimal expected score for the $\ev{post}$ evaluation, based on a scale of 0 to 100. For our context, we defined the optimal expected score for $\ev{post}$ evaluation as 0, as participants should aim to make no mistakes during the second session to maximize their learning gain.

More formally, let $P^A$ and $P^B$ denote the sets of participants in groups $A$ and $B$, respectively. For each participant $p \in \{P^A \cup P^B\}$, two consecutive interviews $I_i(p)$ are conducted, where $i\in{1,2}$. In group $A$, the first interview is conducted using \sys{}. In contrast, in group $B$, the first interview is conducted using \tts{}. Hence, the learning gain of a participant $p \in {P^A}$ associates with \sys{} whereas the learning gain of a participant $p \in {P^B}$ associates with \tts{}. $M_i(p)$ denotes the number of mistakes made by participant $p$ in the interview $I_i(p)$. The learning gain $G(p)$ of participant $p$ is then calculated as follows:
\begin{equation}
\mathit{G(p)} = \dfrac{\mathit{M_2(p)}-\mathit{M_1(p)}}{0-\mathit{M_1(p)}}
\end{equation}  

\paragraph{Processing speed} is a measure of how fast an individual can process information and respond to the surrounding environment~\citep{salthouse1996processing}. It is heavily influenced by the cognitive demands of the task at hand (such as reading, interviewing, or talking), the time limitations associated with it, and the social interactions that the person has to manage while performing the task. We designed an in-context measure of \textit{processing speed} $\mathit{PS}$ to quantify how quickly participants process each turn of the interview~\citep{gorer_roboreit}. It is calculated by dividing the task load $\mathit{TL}$ by the time required to complete the task $\mathit{RT}$ in an interview turn, as shown in Equation~\ref{eq:ps} where $\mathit{TL}$ denotes the task load, and $RT$ represents the response time required to complete the task, and $t$ denotes the interview turn index of participant $p$ trained with system $sys$. We investigate the measure across the two systems, \sys{} and \tts{}, which employ setups with different interaction interfaces that hold distinct levels of social complexity. Indeed, utilizing response time alone as a metric is inadequate due to the different scenarios presented by each system, which directly impact the nature of the task involved. 
\begin{equation}
    \mathit{PS}_{t}(p, sys) = \dfrac{\mathit{TL}_{t}(p, sys)}{\mathit{RT}_{t}(p, sys)}
\label{eq:ps}
\end{equation}

In each interview turn, the participant is presented with three options and expected to select one of them. We measure the response time $\mathit{RT}$ from when the participant is provided with the options until they respond. We calculate the task load per turn as the effort required in evaluating the options. As the options get longer and similar to each other, the participant is expected to put more effort into evaluating them to pick the correct option. The evaluation effort is estimated as the reading effort factored by the difficulty of an interview turn -- which indicates how challenging it is for the participant to select the correct answer from the available options. According to the education literature~\citep{ascalon2007distractor,shin2019multiple}, one of the elements of the difficulty index is the similarity of the multiple-choice options, which makes it harder for the examinee to eliminate incorrect options and identify the differences between the options. We calculate the similarity of the options using Universal Sentence Encoder\footnote{https://tfhub.dev/google/universal-sentence-encoder/4}~\citep{cer2018universal}, which converts the text of each option item to a fixed-length vector representation and computes the cosine similarities between the option item vectors. 
More details of the calculation of \textit{processing speed} are available in our previous work~\citep{gorer_roboreit}. 


The processing speed of participant $p$ for the session conducted with system $sys$ is then given by averaging $\mathit{PS}_t(p, sys)$ over all the turns $\mathit{T}$ as follows:
\begin{equation}
\mathit{PS}(p, sys) = \dfrac{1}{T} \sum_{t \in \{1...|T|\}} {\mathit{PS}_t(p, sys)}
\end{equation}

\paragraph{Turn-specific processing speed} $\mathit{PS}^{\psi}(p)$ calculates a participant's $p$ processing speed $\mathit{PS}$ for the given specific group of interview turns $t^{\psi}$, where $\psi$ denotes the group of the turns, i.e., correctly or incorrectly responded turns. It is given by averaging $\mathit{PS}_t^{\psi}(p, sys)$ over all turns, occurred in both \sys{} system $R$ and \tts{} system $V$, belonging to the specified group, as follows:
\begin{equation}
    \mathit{PS}^{\psi}(p) = \dfrac{1}{\sum_{sys \in \{R, V\}} T^{\psi}(p, sys)} \sum_{sys \in \{R, V\}} \sum^{|T^{\psi}(p, sys)|}_{t} {\mathit{PS}_{t}(p, sys)} 
    \end{equation}

\paragraph{Perceived acceptance} refers to an individual's subjective belief or perception of the level of acceptance and usefulness of a particular technology. We investigated the perceptions of the participant $p$ for the acceptance of the given system $sys$. Upon completion of the training session with each system, the participants are asked to score the acceptance of the underlying system $sys$ by the technology acceptance model questionnaire described in Section~\ref{sec:materials}. Using a 5-point Likert scale, the questionnaire measures the overall acceptance in three variables; attitudes $\mathit{PATT}(p, sys)$, perceived ease-of-use $\mathit{PEU}(p, sys)$, and perceived usefulness $\mathit{PU}(p, sys)$. All variables are integer values ranging from $1$ to $5$, with higher values indicating more positive attitudes, ease-of-use, and usefulness towards the technology, respectively.

\paragraph{Engagement} refers to a state of active and meaningful involvement in a particular activity or interaction. A variety of cognitive, emotional, and behavioral processes are involved in this psychological and social phenomenon. While there has been a significant amount of research on engagement, there is still no consensus on a standard measure of engagement~\citep{salam2022automatic}. This is because engagement is a complex and multifaceted construct that can manifest differently in various contexts and settings. There are many different approaches to measuring engagement, and the choice of measure can depend on a range of factors, including the purpose of the assessment, the specific context or activity being studied, and the characteristics of the individuals being assessed. Some commonly used measures of engagement include self-report questionnaires, behavioral observations, and physiological measures. In our study, we utilize both a self-report questionnaire and an analysis of voice signals to measure engagement.
\begin{itemize}
    \item{\textit{Perceived engagement}} score $\mathit{PE}(p, sys)$ is given by participant $p$ by the post-condition questionnaire prepared for the design evaluation of system $sys$. The variable $\mathit{PE}(p, sys)$ takes values in {1,...,5}, where higher values indicate higher engagement.
    \item{\textit{Measured engagement}} score $\mathit{ME}(p, sys)$ is calculated as the average arousal level of participant $p$ during the interview conducted by the system $sys$. Arousal is one of the important indicators of engagement~\citep{ferrari2021using}, and it refers to the level of physiological and psychological activation or stimulation that an individual experiences in response to a particular stimulus or situation. High arousal is often associated with increased engagement and task adaptation in the context of technology use~\citep{beaudry2010other}. 
    Voice is an important modality for emotion transfer, and prosodic features, in particular, have been shown to be strongly correlated with arousal levels~\citep{schirmer2017emotion}. \citet{wagner2022dawn} propose a transformer-based deep neural network model which constitutes the state-of-art in speech emotion recognition. By using their pre-trained model, we automatically extract arousal levels from speech segments. The overall arousal level across an interview is then calculated by taking the mean of the arousal levels of all turns.
\end{itemize}

\subsection{Study Execution}
\label{sec:study_execution}
We introduced the experiment to the students of the ``Software Requirements Engineering” course in the Computer Engineering Department of Bogazici University. The department offers a one-year non-thesis M.Sc. degree for industry professionals besides a regular M.Sc. program in software engineering. This course is a selective graduate course offered in both programs. In order to allow the students to gauge their self-interest in participating in the study, we explained the main goal of the study, eligibility requirements, how the study will be conducted, and how long it will take. We did not mention the details regarding the experiment's research questions or the content of the scenarios used in the experiments to avoid any potential bias. Although the lecturer encouraged the students to take part in the study as an extracurricular class activity, participation was entirely up to the students. The participating students were offered to receive a small bonus grade. 27 students accepted to take part in the study. The study was conducted in the second semester of the 2021-2022 academic year. 

\renewcommand{\arraystretch}{1.2}
\begin{table*}
\scriptsize
\caption{Demographics data of the participants provided separately for \textit{Group A} and \textit{Group B} as well as for all participants.}
\label{tab:participant_demographics}
\begin{tabular}{ll|cc|c}
\hline
                                                                                                                   & \textit{}                      & \textbf{\begin{tabular}[c]{@{}c@{}}Group A\\ n=13\end{tabular}} & \textbf{\begin{tabular}[c]{@{}c@{}}Group B\\ n=14\end{tabular}} & \textbf{\begin{tabular}[c]{@{}c@{}}All\\ n=27\end{tabular}} \\ \hline
\multirow{2}{*}{Age(years)}                                                                                        & \textit{range}                 & 23 - 36                                                         & 24 - 42                                                         & 23-42                                                       \\
                                                                                                                   & \textit{mean (SD)}             & 26.8 (3.8)                                                      & 29.3 (5.1)                                                      & 28.1 (4.7)                                                  \\ \hline
\multirow{2}{*}{Gender}                                                                                            & \textit{Female}                & 6                                                               & 3                                                               & 9                                                           \\
                                                                                                                   & \textit{Male}                  & 7                                                               & 11                                                              & 18                                                          \\ \hline
\multirow{2}{*}{Occupation}                                                                                        & \textit{SW professsional}      & 9                                                               & 12                                                              & 21                                                          \\
                                                                                                                   & \textit{Other}                 & 4                                                               & 2                                                               & 6                                                           \\ \hline
\multirow{4}{*}{\begin{tabular}[c]{@{}l@{}}Years of Work \\ Experience\end{tabular}}                               & \textit{0-1 years:}            & 0                                                               & 1                                                               & 1                                                           \\
                                                                                                                   & \textit{1-3 years:}            & 10                                                              & 7                                                               & 18                                                          \\
                                                                                                                   & \textit{4-6 years:}            & 1                                                               & 3                                                               & 4                                                           \\
                                                                                                                   & \textit{\textgreater 6 years:} & 2                                                               & 3                                                               & 5                                                           \\ \hline
\begin{tabular}[c]{@{}l@{}}Confidence level in \\ practicing RE interviews\\ (1-lowest, 5-highest)\end{tabular}    & \textit{mean (SD)}             & 2.84 (0.77)                                                     & 3.07 (0.70)                                                     & 2.96 (0.74)                                                 \\ \hline
\begin{tabular}[c]{@{}l@{}}Level of theoretical \\ knowledge on RE interviews\\ (1-lowest, 5-highest)\end{tabular} & \textit{mean (SD)}             & 3.00 (0.87)                                                     & 3.21 (0.77)                                                     & 3.11 (0.83)                                                 \\ \hline
\multirow{4}{*}{\# of practiced RE interviews}                                                                     & \textit{0}                     & 6                                                               & 6                                                               & 12                                                          \\
                                                                                                                   & \textit{1-3 times}             & 5                                                               & 6                                                               & 11                                                          \\
                                                                                                                   & \textit{4-6 times}             & 1                                                               & 0                                                               & 1                                                           \\
                                                                                                                   & \textit{\textgreater 6 times}  & 1                                                               & 2                                                               & 3                                                           \\ \hline
\multirow{4}{*}{\begin{tabular}[c]{@{}l@{}}\# of real RE interviews \\ conducted\end{tabular}}                     & \textit{0}                     & 10                                                              & 6                                                               & 16                                                          \\
                                                                                                                   & \textit{1-3 times}             & 2                                                               & 5                                                               & 7                                                           \\
                                                                                                                   & \textit{4-6 times}             & 0                                                               & 0                                                               & 0                                                           \\
                                                                                                                   & \textit{\textgreater 6 times}  & 1                                                               & 3                                                               & 4                                                           \\ \hline
\begin{tabular}[c]{@{}l@{}}Level of Interview Anxiety\\ (1-lowest, 5-highest)\end{tabular}                         & \textit{mean (SD)}             & \multicolumn{1}{l}{2.46 (0.92)}                                 & \multicolumn{1}{l|}{1.92 (0.79)}                                & 2.18 (0.90)                                                 \\ \hline
\begin{tabular}[c]{@{}l@{}}Level of Negative Attitude \\ Towards Robots \\ (1-lowest, 5-highest)\end{tabular}      & \textit{mean (SD)}             & \multicolumn{1}{l}{2.30 (0.74)}                                 & \multicolumn{1}{l|}{2.39 (0.80)}                                & 2.35 (0.78)                                                 \\ \hline
\end{tabular}
\end{table*}
\renewcommand{\arraystretch}{1}

The participants, 9 females and 18 males ranging from 23 to 42 years of age ($\mathit{mean}=28.1, \mathit{SD}=4.7$), were all graduate students of the software engineering master program. We gathered information about their professions and levels of seniority at work. They all had full-time jobs. Out of the 27, 21 were employed in the software industry as developers, testers, analysts, or similar roles. The remaining six were working in fields other than software engineering, such as mechanical engineering  and portfolio management. The majority of participants hold entry-level roles in their jobs and have worked for less than a year $(n=1)$, between 1-3 years $(n=18)$, between 4-6 years $(n=4)$, or longer than 6 years $(n=5)$.

We assessed the participants' knowledge of and experience with software requirements elicitation interviews. All of them were receiving training in software requirements engineering through the university course, and three of them had also joined company training programs related to this topic before. 15 of them practiced mock interviews before, one to three times $(n=11)$, four to six times $(n=1)$, and more than six times $(n=3)$. 12 of them had never participated in a mock interview. On the other hand, the number of participants who conducted a real interview is relatively lower. Only 11 of them had an interview with a real stakeholder one to three times $(n=7)$, more than six times $(n=4)$. We also gathered participants' self-evaluation scores on theoretical understanding of requirement elicitation interview techniques ($\mathit{mean}=3.11, \mathit{SD}=0.83$) and confidence level in practicing an interview ($\mathit{mean}=2.96, \mathit{SD}=0.74$) over a 5-point Likert scale (1=poor, 5=very good). The interview anxiety level is measured on a scale of 1 to 5, where 1 represents the lowest level of anxiety and 5 represents the highest level of anxiety. The mean anxiety level of the participants was 2.18 ($\mathit{SD}=0.90$). The participants' negative attitude score towards robots was 2.35 on average ($\mathit{SD}=0.78$). It is also measured on a scale of 1 to 5, where 1 represents the lowest level of negative attitudes toward robots and 5 represents the highest level of negative attitudes towards robots. None of the participants were native English speakers, but they were all proficient in the language with varying levels. The participants' demographics are given in Table~\ref{tab:participant_demographics}.

Out of the 27 participants, 13 are assigned to \textit{Setup A} condition (Group A), and the other 14 are assigned to \textit{Setup B} condition (Group B) in a randomized manner. Using an online scheduling tool, the experiment hours were set in advance based on the participants' preferences. At the appointed time, the participant and experimenter joined the call, and the experiment was carried out using the procedures outlined in Section~\ref{sec:procedure}. The interview session of the experiment lasted 11.81 minutes on average ($\mathit{SD}=2.71$) for \sys{} condition and 11.22 minutes ($\mathit{SD}=2.34$) for \tts{} condition. The difference is mostly because of the slightly slower speech rate of the text-to-speech module utilized in \sys{} compared to the one used in \tts{}. The interview session lengths show similar deviations across the participants for both conditions, which is caused by the participant's processing speed and how many turns they have. The participants also spend considerable time in the feedback session with an average of 9.44 minutes ($\mathit{SD}=2.37$) in \sys{} condition and 6.98 minutes ($\mathit{SD}=1.93$) in \tts{} condition. The number of mistakes visited and the participants' review time for their second attempts account for most of the differences in feedback session duration among the participants and between conditions. The overall study was completed in 20 days period.

\subsubsection{Randomization Validation}
\label{sec:randval}
We checked the validation of randomization to ensure that there is no systematic bias in the way participants are allocated to \textit{Group A} and \textit{Group B}. As the participants' baseline characteristics, we monitored their self-reported NARS scores, interview anxiety scores, and expertise levels in requirements elicitation interviews across the two groups. We believe that the unequal distribution of these elements could influence the results of the related research objectives. We applied Mann-Whithey U test~\citep{nachar2008mann} to check if there was a significant difference between the two groups as the data for the three factors are not normally distributed. The participants assigned to the \textit{Group A} and \textit{Group B} did not significantly differ in their NARS scores ($U=87.0, \textit{p-value}=0.85$) and interview anxiety scores ($U=121.0, \textit{p-value}=0.13$). Likewise, the two groups' expertise levels in conducting requirements elicitation interviews did not show a significant discrepancy. The average of the self-reported theoretical and practical expertise scores was considered as the expertise level, yet there was no difference ($U=83.0, \textit{p-value}=0.70$).

To maintain a consistent level of difficulty and interview length, the scenarios employed in the experiment are meticulously designed and randomized across the experimental conditions. We checked if the associated scenarios influence the evaluation of the systems for each dependent variable. A series of statistical tests show that there is not any significant effect of the scenario on the results. 

Overall, our analysis reveals that there are no deviations from the intended randomization scheme. 

\subsection{Results}
\label{sec:results}

We conduct a series of statistical tests to examine a set of hypotheses derived from our research questions (RQs). The null and alternative hypotheses corresponding to each RQ are presented, along with the specific tests used to analyze them. We first perform Shapiro-Wilk normality tests to determine if the data is are normally distributed. For dependent samples of data that are not normally distributed, we employ the Wilcoxon signed-rank test~\citep{woolson2007wilcoxon}. For the data that fulfill the characteristics of normal distribution, we use independent $T$-test to compare the means of two independent groups and dependent (paired) $T$-test to compare the means of two measurements taken from the same participants~\citep{cramer2016mathematical}. All hypotheses are tested at a 95\% confidence level ($\textit{p-value} \leq 0.05$). 

For our study, Likert scales are ordinal, with 1 denoting a strong disagreement and 5 denoting a strong agreement. It is not safe to assume that the intervals between the Likert values are the same even though they are ranked in a certain order. Because the mathematical processes required to obtain the mean and standard deviation are inappropriate for ordinal data, the common approach is to use the median as the measure of central tendency~\citep{blaikie2003analyzing}. We use Wilcoxon signed-rank test to evaluate our hypotheses about the variables that are assessed using a Likert scale as it has similar power to the $T$-test even for small sample sizes~\citep{de2010five}.

The descriptive statistics of the dependent variables of each question are reported in the related research question, including median ($\mathit{Mdn}$) and interquartile range ($\mathit{IQR}$) for the data that are not normally distributed, and mean ($\mathit{Mean}$) and standard deviation ($\mathit{SD}$) for normally distributed data. These statistics provide information on the central tendency and dispersion of the data, and allow for a comparison of the groups being studied. The corresponding hypothesis test results are also presented with $\textit{p-value}$ and test statistics.

\rqdef{RQ1: }\textit{How do \sys{} and \tts{} influence the learning gain of the participants?}\\
To answer RQ1, we used independent samples of the learning gains of the participants of Group A and Group B. The definition of learning gain measurement is given in Section~\ref{sec:dependent_vars}. In Group A, the participants trained with \sys{} first, followed by \tts{}, whereas in Group B, it is the opposite. The participant's learning gain is associated with the system utilized during the initial session, given that the learning impact of the system is measured by the difference in success across consecutive interview sessions. We argue that the learning gains of the participants trained with \sys{} in their first sessions are different from the learning gains of the participants who trained with \tts{} in their first sessions. Formally we have $G_\sys{} = \{G(p_i), i=1 ... |P^A|\}$ and $G_\tts{} = \{G(p_i), i=1 ... |P^B|\}$. The two-tailed null hypothesis is $H_{10} = $``the participants' learning gain who are trained with \sys{} is equal to the ones who are trained with \tts{}" (i.e., $\mu_{G_\sys{}} = \mu_{G_\tts{}}$). The two-tailed alternative hypothesis is $H_{11} = $``the participants' learning gain who are trained with \sys{} is not equal to the ones who are trained with \tts{}" (i.e., $\mu_{G_\sys{}} \neq \mu_{G_\tts{}}$). Two-tailed independent T-test reveals significant results ($T=2.01, \textit{p-value}=0.05$). Hence, we can reject $H_{10}$ in favour of $H_{11}$. The learning gain of the participants trained with \sys{} is higher than those trained with \tts{}. The test result and descriptive statistics are given in Table~\ref{tab:dep_var_rq1}.

\begin{table*}
    \caption{The descriptive statistics and hypothesis test results for RQ1.}
    \label{tab:dep_var_rq1}
    \begin{tabular}{p{3.75cm}R{1cm}R{1cm}R{1cm}R{1cm}R{3.75cm}}
    \hline
                                & \multicolumn{2}{c|}{Group A}                     & \multicolumn{2}{c|}{Group B}                     & \multicolumn{1}{r}{\begin{tabular}[r]{@{}r@{}}Independent T-test\end{tabular}}         \\ \hline
    \multicolumn{1}{r}{}        & $\mathit{Mean}$                 & \multicolumn{1}{r|}{$\mathit{SD}$}   & $\mathit{Mean}$                 & \multicolumn{1}{r|}{$\mathit{SD}$}   & p-value (T stat)                                                                         \\ \hline
    Learning Gain (G)           & 0.35                 & \multicolumn{1}{r|}{0.28} & 0.08                 & \multicolumn{1}{r|}{0.39} & \textbf{0.05 (2.01)}                                                                     \\ \hline
    \end{tabular}
    \end{table*}
\renewcommand{\arraystretch}{1.0}

\rqdef{RQ2: }\textit{How do \sys{} and \tts{} influence the processing speed of the users during the interview session?}\\
We consider dependent samples of the processing speed variable $\mathit{PS}$ from the same participant experimented with the two systems. Formally, we have $\mathit{PS}_\sys{} = \{\mathit{PS(p_i, \sys{})}$, $i=1 ... |P|\}$ and $\mathit{PS}_\tts{} = \{\mathit{PS(p_i, \tts{})}, i=1 ... |\mathit{P}|\}$. The two-tailed null hypothesis is $H_{20} = $``the processing speed in \sys{} condition is equal to the one of the \tts{} condition" (i.e., $\mu_{\mathit{PS}_\sys{}} = \mu_{\mathit{PS}_\tts{}}$). The two-tailed alternative hypothesis is $H_{21} = $``the processing speed in \sys{} condition is not equal to the one of the \tts{} condition" (i.e., $\mu_{\mathit{PS}_\sys{}} \ne \mu_{\mathit{PS}_\tts{}}$). Since the processing speed variable violates the normal distribution assumption with Shapiro-Wilk's test result ($\mathit{W}=0.89, \textit{p-value} = 0.01$), we applied Wilcoxon signed-rank test to check whether the processing speed is significantly different in \sys{} and \tts{} conditions ($H_{21}$). Although the processing speed is greater in \sys{} condition, the difference is not significant with $\mathit{Z}=181.0, \textit{p-value}=0.86$ (as given in Table~\ref{tab:dep_var_rq2}). Hence, we can not reject $H_{20}$.

\begin{table*}
    \caption{The descriptive statistics and hypothesis test results for RQ2.}
    \label{tab:dep_var_rq2}
    \begin{tabular}{p{3.75cm}R{1cm}R{1cm}R{1cm}R{1cm}R{3.75cm}}
    \hline
                                & \multicolumn{2}{c|}{\sys{}}                      & \multicolumn{2}{c|}{\tts{}}                      & \multicolumn{1}{r}{\begin{tabular}[r]{@{}r@{}}Wilcoxon signed  rank test\end{tabular}} \\ \hline
                                & $\mathit{Mdn}$               & \multicolumn{1}{r|}{$\mathit{IQR}$}  & $\mathit{Mdn}$               & \multicolumn{1}{r|}{$\mathit{IQR}$}  & p-value (Z stat)                                                                         \\ \hline
    Processing Speed (PS)       & 7.11                 & \multicolumn{1}{l|}{4.36} & 8.06                 & \multicolumn{1}{c|}{3.05} & 0.86 (181.0)                                                                             \\ \hline
    \end{tabular}
    \end{table*}
\renewcommand{\arraystretch}{1.0}
    
\rqdef{RQ3: }\textit{Does the participants' processing speed for a question vary based on their performance on the question?}\\
To answer this RQ, we compared the processing speed of the participants on turns with no mistake that they responded with a correct option versus the mistaken turns, and see if there is a significant difference between the two. Our goal is to examine whether the processing speed $\mathit{PS}$ of the participant is influenced by the accuracy of the answer.
We consider dependent samples of the turn-specific processing speed variable $\mathit{PS^{\psi}(p)}$ for no-mistake ($\mathit{\psi=NM}$) and mistaken ($\mathit{\psi=M}$) responses from each participant. The two-tailed null hypothesis is defined as $H_{30} = $``The processing speed for the responses with \textit{no-mistake} is equal to the responses with \textit{mistake}." (i.e., $\mathit{PS^{NM}(p) = PS^{M}(p)}$). The two-tailed alternative hypothesis is $H_{31} = $``the processing speed for the responses with \textit{no-mistake} is not equal to the ones with \textit{mistake}" (i.e., $\mathit{PS^{NM}(p) \neq PS^{M}(p)}$). To test the hypothesis, we performed two-tailed Wilcoxon signed-rank test. The difference is significant with ($\mathit{Z}=325.0, \textit{p-value}<0.001$) and we rejected $H_{30}$. The descriptive statistics and test results are given in Table~\ref{tab:dep_var_rq3}. Participants processed questions they correctly answered more quickly than questions they incorrectly answered.

\begin{table*}
    \caption{The descriptive statistics and hypothesis test results for RQ3.}
    \label{tab:dep_var_rq3}
    \begin{tabular}{p{3.75cm}R{1cm}R{1cm}R{1cm}R{1cm}R{3.75cm}}
    \hline
                                & \multicolumn{2}{c|}{Mistake}                     & \multicolumn{2}{c|}{No Mistake}                  & \multicolumn{1}{c}{\begin{tabular}[c]{@{}c@{}}Wilcoxon signed rank test\end{tabular}} \\ \hline
                                & $\mathit{Mdn}$               & \multicolumn{1}{r|}{$\mathit{IQR}$}  & $\mathit{Mdn}$               & \multicolumn{1}{r|}{$\mathit{IQR}$}  & p-value (Z stat)                                                                         \\ \hline
    Processing Speed (PS)       & 7.34                 & \multicolumn{1}{r|}{2.82} & 8.69                 & \multicolumn{1}{r|}{3.62} & \textbf{\textless 0.001 (325.0)}                                                         \\ \hline
                               
    \end{tabular}
    \end{table*}
\renewcommand{\arraystretch}{1.0}
    
\rqdef{RQ4: }\textit{How do \sys{} and \tts{} influence the perceived acceptance of the underlying system in the dimensions of \textbf{RQ4a:} perceived attitudes, \textbf{RQ4b:} perceived ease-of-use, \textbf{RQ4c:} perceived usefulness}?\\
To answer the sub-research questions RQ4a, RQ4b, and RQ4c, we consider dependent samples of the dependent variables $\mathit{PATT}$, $\mathit{PEU}$, and $\mathit{PU}$, respectively, from condition \sys{} and \tts{}. To evaluate the reliability of the questions in each of the surveyed dimensions, we employed Cronbach’s $\alpha$ analysis, a widely used method for assessing the internal consistency of data. %
The reliability of the questions in each of the surveyed dimensions was higher than 0.70, indicating high reliability, particularly the questions of the dimensions of perceived attitudes towards using ($\alpha=0.87$), perceived ease-of-use ($\alpha=0.89$), and perceived usefulness ($\alpha=0.87$). The scores of each dimension's questions were added up and averaged for each participant in each experimental condition. The average point of each dimension was then used to run the significance tests and to calculate descriptive statistics, which are given in Table~\ref{tab:dep_var_rq4}.
    
\begin{figure*}
    \centering
    \includegraphics[scale=0.6]{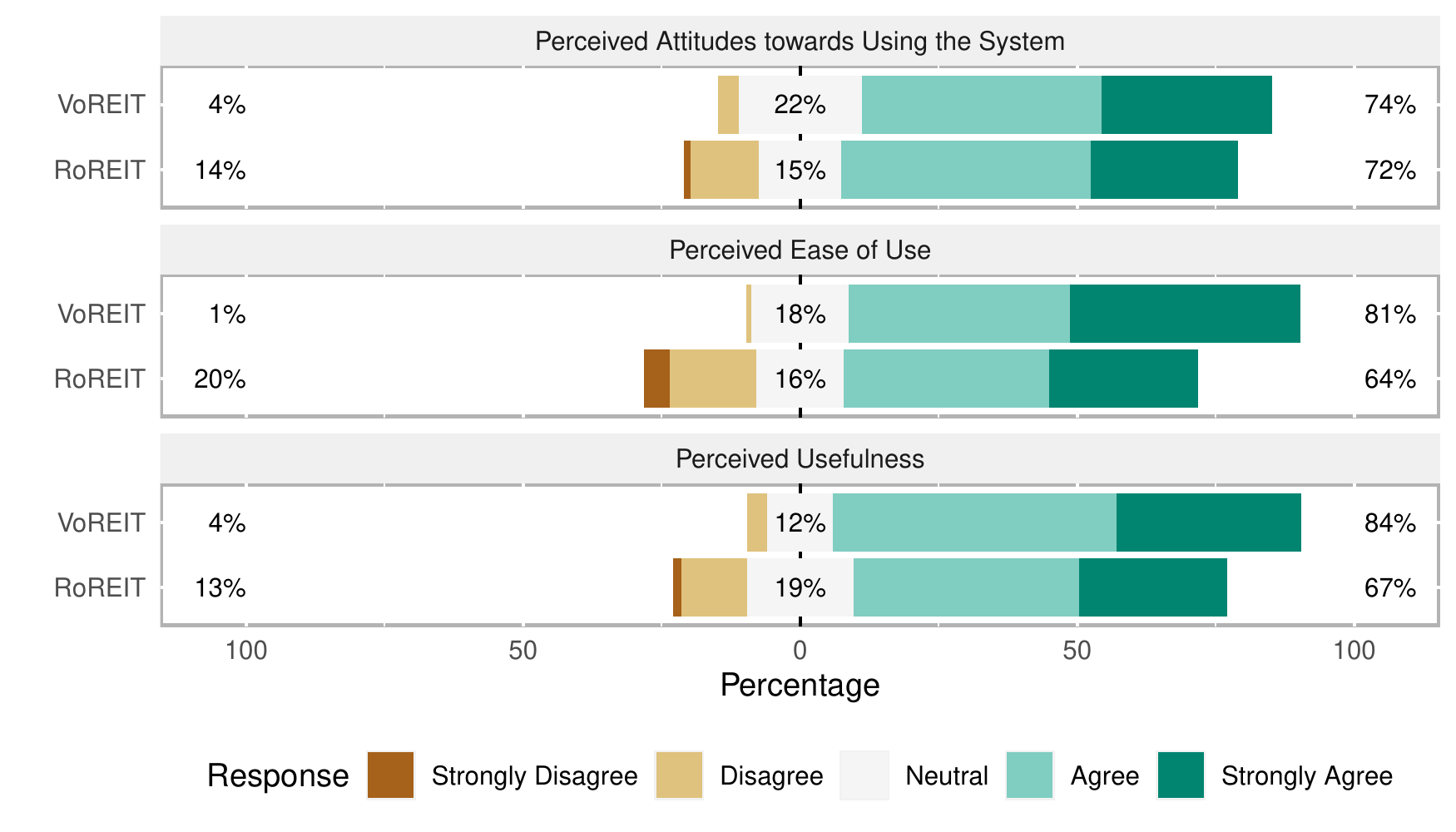}
    \caption{The questionnaire results for the technology acceptance model for the conditions of \tts{} and \sys{} in the dimensions of attitudes toward using, ease-of-use, and usefulness.}
    \label{fig:p_attitudes}
    \end{figure*}
    
\rqdef{RQ4a: }We have $\mathit{PATT}_\sys{} = \{\mathit{PATT}(p_i, \sys{}), i=1 ... |\mathit{P}|\}$ and $\mathit{PATT}_\tts{} = \{\mathit{PATT}(p_i, \tts{}), i=1 ... |\mathit{P}|\}$. The two-tailed null hypothesis is $H_{4a0} = $``the perceived attitudes in \sys{} condition is equal to the one of the \tts{} condition" (i.e., $\mu_{\mathit{PATT}_\sys{}} = \mu_{\mathit{PATT}_\tts{}}$). The two-tailed alternative hypothesis is $H_{4a1} = $``the perceived attitudes for \sys{} condition is not equal to the one for the \tts{} condition" (i.e., $\mu_{\mathit{PATT}_\sys{}} \ne \mu_{\mathit{PATT}_\tts{}}$). We applied two-tailed Wilcoxon signed-rank test to test the significance. The difference is not significant ($\mathit{Z}=59.5, \textit{p-value}=0.41$); hence, we failed to reject $H_{4a0}$. From Figure \ref{fig:p_attitudes}, we see that a high proportion of both groups evaluated the interview experience positively. 
    
\rqdef{RQ4b: }Formally, we have $\mathit{PEU}_\sys{} = \{\mathit{PEU}(p_i, \sys{}), i=1 ... |\mathit{P}|\}$ and $\mathit{PEU}_\tts{} = \{\mathit{PEU}(p_i, \tts{}), i=1 ... |\mathit{P}|\}$. The two-tailed null hypothesis is $H_{4b0} = $``the perceived ease-of-use in \sys{} condition is greater than equal to the one of the \tts{} condition" (i.e., $\mu_{\mathit{PEU}_\sys{}} = \mu_{\mathit{PEU}_\tts{}}$). The two-tailed alternative hypothesis is $H_{4b1} = $``the perceived ease-of-use for \sys{} condition is not equal the one for the \tts{} condition" (i.e., $\mu_{\mathit{PEU}_\sys{}} \neq \mu_{\mathit{PEU}_\tts{}}$). Two-tailed Wilcoxon signed-rank test result reveals a significant difference ($\mathit{Z}=40.0, \textit{p-value}=0.03$) between the two groups for perceived ease-of-use. As seen from Figure \ref{fig:p_attitudes}, the ratio of responses on the strong agreement of ease-of-use is higher in \tts{} condition compared to \sys{}. The results indicate that \tts{} is perceived to be easier to use compared to \sys{} though both systems are rated highly positive.
    
\rqdef{RQ4c: }Formally, we have $\mathit{PU}_\sys{} = \{\mathit{PU}(p_i, \sys{}), i=1 ... |\mathit{P}|\}$ and $\mathit{PU}_\tts{} = \{\mathit{PU}(p_i, \tts{}), i=1 ... |\mathit{P}|\}$. The two-tailed null hypothesis is $H_{4c0} = $``the perceived usefulness in \sys{} condition is equal to the one of the \tts{} condition" (i.e., $\mu_{\mathit{PU}_\sys{}} = \mu_{\mathit{PU}_\tts{}}$). The two-tailed alternative hypothesis is $H_{4c1} = $``the perceived usefulness for \sys{} condition is not equal to the one for the \tts{} condition" (i.e., $\mu_{\mathit{PU}_\sys{}} \ne \mu_{\mathit{PU}_\tts{}}$). Two-tailed Wilcoxon signed-rank test shows the difference is not significant for the perceived usefulness for \sys{} and \tts{} ($\mathit{Z}=13.5, \textit{p-value}=0.13$). As shown in Figure \ref{fig:p_attitudes}, participants evaluated both systems as highly useful.

\begin{table*}
    \caption{The descriptive statistics and hypothesis test results for RQ4.}
    \label{tab:dep_var_rq4}
    \begin{tabular}{p{4.25cm}R{1cm}R{1cm}R{1cm}R{1cm}R{3.5cm}}
    \hline
                                & \multicolumn{2}{c|}{\sys{}}                      & \multicolumn{2}{c|}{\tts{}}                      & \multicolumn{1}{r}{\begin{tabular}[c]{@{}c@{}}Wilcoxon signed rank test\end{tabular}} \\ \hline
                                & $\mathit{Mdn}$               & \multicolumn{1}{r|}{$\mathit{IQR}$}  & $\mathit{Mdn}$               & \multicolumn{1}{r|}{$\mathit{IQR}$}  & p-value (Z stat)                                                                         \\ \hline
    Perceived Attitudes ($\mathit{PATT}$)  & 4                    & \multicolumn{1}{r|}{0.75} & 4                    & \multicolumn{1}{r|}{0.75} & 0.41 (59.5)                                                                              \\
    Perceived Ease of Use ($\mathit{PEU}$) & 4                    & \multicolumn{1}{r|}{1.5}  & 4                    & \multicolumn{1}{r|}{1}    & \textbf{0.03 (40.0)}                                                                     \\
    Perceived Usefulness ($\mathit{PU}$)   & 4                    & \multicolumn{1}{r|}{2}    & 4                    & \multicolumn{1}{r|}{1}    & 0.13 (13.5)                                                                       \\ \hline
    \end{tabular}
    \end{table*}
\renewcommand{\arraystretch}{1.0}
        
\rqdef{RQ5: }\textit{How do \tts{} and \sys{} influence the perceived and measured engagement levels during training with the system?}\\
To answer RQ5, we considered both the perceived engagement levels reported by the participants and objectively measured engagement levels from the voice signals of the participants. For the first, we used dependent samples of the perceived engagement scores $\mathit{PE}$ provided for both systems. Formally, we have $\mathit{PE}_\sys{} = \{\mathit{PE}(p_i, \sys{}), i=1 ... |\mathit{P}|\}$ and $\mathit{PE}_\tts{} = \{\mathit{PE}(p_i, \tts{}), i=1 ... |P|\}$. The two-tailed null hypothesis is $H_{510} = $``the perceived engagement in \sys{} condition is equal to the one of the \tts{} condition" (i.e., $\mu_{\mathit{PE}_\sys{}} = \mu_{\mathit{PE}_\tts{}}$). The two-tailed alternative hypothesis is $H_{511} = $``the perceived engagement for \sys{} condition is not equal to the one for the \tts{} condition" (i.e., $\mu_{\mathit{PE}_\sys{}} \neq \mu_{\mathit{PE}_\tts{}}$). The result shows significance ($\mathit{Z}=39.0, \textit{p-value}=0.04$) so that $H_{510}$ is rejected. The participants evaluated \tts{} as more engaging than \sys{}. As the second hypothesis, we postulate that there is no significant difference in the mean arousal level of the participants' speech across the two systems. Formally, we have $\mathit{ME}_\sys{} = \{\mathit{ME}(p_i, \sys{}), i=1 ... |\mathit{P}|\}$ and $\mathit{ME}_\tts{} = \{\mathit{ME}(p_i, \tts{}), i=1 ... |\mathit{P}|\}$. The two-tailed null hypothesis is $H_{520} = $``the measured engagement level in \sys{} condition is equal to the one of the \tts{} condition" (i.e., $\mu_{\mathit{ME}_\sys{}} = \mu_{\mathit{ME}_\sys{}}$). The two-tailed alternative hypothesis is $H_{521} = $``the measured engagement for \sys{} condition is not equal to the one for the \tts{} condition" (i.e., $\mu_{\mathit{ME}_\sys{}} \ne \mu_{\mathit{ME}_\tts{}}$). The relative $T$-test result shows no significance ($\mathit{T}=0.45, \textit{p-value}=0.65$) so that $H_{520}$ can not be rejected. The perceived engagement scores are not equated with the measured levels from the voice signals of the participants. The statistics and test results for both dependent variables are provided in Table~\ref{tab:dep_var_rq5}.

\begin{table*}
    \caption{The descriptive statistics and hypothesis test results for RQ5.}
    \label{tab:dep_var_rq5}
    \begin{tabular}{p{4.25cm}R{1cm}R{1cm}R{1cm}R{1cm}R{3.5cm}}
    \hline
                                & \multicolumn{2}{c|}{\sys{}}                      & \multicolumn{2}{c|}{\tts{}}                      & \multicolumn{1}{r}{\begin{tabular}[c]{@{}c@{}}Wilcoxon signed rank test\end{tabular}} \\ \hline
                                & $\mathit{Mdn}$               & \multicolumn{1}{r|}{$\mathit{IQR}$}  & $\mathit{Mdn}$               & \multicolumn{1}{r|}{$\mathit{IQR}$}  & p-value (Z stat)                                                                         \\ \hline
    Perceived Engagement ($\mathit{PE}$)   & 4                    & \multicolumn{1}{r|}{1}    & 4                    & \multicolumn{1}{r|}{0}    & \textbf{0.02 (39.0)}                                                                              \\ \hline \\ \hline
                                & \multicolumn{2}{c|}{\sys{}}                      & \multicolumn{2}{c|}{\tts{}}                      & \multicolumn{1}{r}{\begin{tabular}[c]{@{}c@{}}Dependent T-test\end{tabular}}          \\ \hline
    \multicolumn{1}{r}{}        & $\mathit{Mean}$                 & \multicolumn{1}{r|}{$\mathit{SD}$}   & $\mathit{Mean}$                 & \multicolumn{1}{r|}{$\mathit{SD}$}   & p-value (T stat)                                                                         \\ \hline
    Measured Engagement ($\mathit{ME}$)    & 0.46                 & \multicolumn{1}{r|}{0.09} & 0.45                 & \multicolumn{1}{r|}{0.09} & 0.65 (0.45)                                                                              \\ \hline
    \end{tabular}
    \end{table*}
\renewcommand{\arraystretch}{1.0}
    
\rqdef{RQ6: }\textit{What are the relationships between individual user characteristics (i.e., age, gender, interview experience level, interview anxiety level) and perceived acceptance and engagement of the systems?}\\
For each of the user characteristics, we calculated its correlation with the dependent variables, namely perceived attitudes, perceived ease of use, perceived usefulness, and perceived engagement. This analysis is performed for both \sys{} and \tts{} as the dependent variables are collected separately for both of the systems. We use Kendall's rank correlation coefficient $\tau$, which is a statistic used to measure the ordinal association between the two variables. In total, 32 analyses were conducted, and four of them showed statistically significant correlations. The analysis revealed that age has a significant weak positive correlation with perceived attitudes towards using the system for \tts{} ($\tau=0.35, \textit{p-value}=0.02$). Moreover, the correlation between interview anxiety level and perceived usefulness indicated a significant weak negative correlation for \tts{} ($\tau=-0.35, \textit{p-value}=0.04$). The interview anxiety level was found to have a significant moderate negative correlation with perceived engagement for \tts{}($\tau=-0.47, \textit{p-value}=0.007$). For the \sys{} system, the only significant correlation was observed between interview experience level and perceived ease of use with a moderate negative correlation ($\tau=-0.41, \textit{p-value}=0.01$). These findings provide valuable insights into the relationship between user characteristics and system perceptions, which can inform the design of future personalized interview training systems. Further discussions are provided in Section~\ref{sec:discussions}.

\section{Discussion}
\label{sec:discussions}

Our evaluation of the effect of \sys{} and \tts{} on the learning gain of the participants suggests that the physically embodied robotic agent of \sys{} resulted in higher learning gains than the disembodied voice agent of \tts{}. This result is consistent with the literature hypothesizing that embodied pedagogical agents have high capabilities to mimic real-world learning conditions and immerse students in learning activities~\citep{grivokostopoulou2020effectiveness,davis2023meta}. However, existing studies that evaluate embodied trainer agents in empirical research have inconclusive or inconsistent results for this claim~\citep{darwish2014persona}. Further research is needed to better understand the underlying processes and factors affecting the learning gain.

Participants demonstrated quicker processing times for questions they answered correctly compared to those they answered incorrectly. Leveraging our in-context processing speed measure, we can pinpoint questions that participants find challenging and provide timely assistance during the interview. This assistance might entail offering additional explanations or context to help learners in arriving at the correct answer.

\tts{} is perceived to be easier to use compared to \sys{}. Several factors may contribute to this. One would consider the voice-only interaction modality of \tts{} to be less challenging than the audio-visual interaction modality of \sys{}, which requires the participants to keep their cameras on. While physical robots may offer a higher sense of social presence, they may also create pressure for users to perform well and interact in a certain way. In contrast, a voice-only agent may feel more anonymous and less intimidating, allowing users to interact more freely. Familiarity with the underlying technology could also be another important factor. 
People may be unpracticed to interact with physical robots, which can raise their perceived complexity and decrease user comfort. The participants might also think that \tts{} is more accessible and manageable compared to \sys{}, which has limited serving capabilities due to the physical robot and, therefore, is more difficult to use. 

Participants have reported a lower level of engagement with \sys{} compared to \tts{}, even though both systems received high engagement ratings. This discrepancy might be influenced by the relatively slower speech rate in \sys{} as slower speech may result in reduced user involvement and interest. Furthermore, the presence effect, as supported by prior research~\citep{li2015benefit}, is another plausible explanation for this result. They show that users tend to hold more favorable views of physical robots when they are physically present in their environment, as opposed to when they are represented only through digital means, such as video feeds on a screen. Additionally, individual preferences and biases could have played a role in shaping participants' perceptions of engagement with the two systems. Although the measured engagement levels did not reveal a preference for \tts{} over \sys{}, there remains potential for improving \sys{}'s perceived engagement. This could be achieved by enhancing the robot's voice for more effective communication or by incorporating the robot's physical presence into the user interaction.

Age has been known to play an important role in shaping one's perspective and adoption of new technologies~\citep{hong2013old}. Older adults often exhibit a greater sense of comfort when interacting with familiar technologies, in contrast to advanced technologies with which they have limited exposure.
Our empirical data also reveal a positive correlation between the age of the participants and the perceived attitude towards \tts{}. Additionally, our analysis demonstrates a negative correlation between participants' levels of interview experience and their perceived ease of use of \sys{}. Since the participants with more interview experience are more likely to have their own preferences for how interviews should be conducted, their expectations from a robotic training system could be more higher. In cases where the system falls short of meeting these expectations, it can lead to a perception of reduced ease of use, as they may compare the system's performance to their own established standards and practices.

Anxiety can make people self-conscious and focus on how well their performance is perceived by the interviewer rather than on the process itself, which may exacerbate the interview experience for both sides. Previous research on the use of technological tools for job interviews claims that a person's anxiety may be induced by several variables, such as the setting of the interview and the realism of the technology involved~\citep{kwon2013level,vilar2020can}. %
Our analysis reveals that the participants' levels of anxiety negatively correlate with the perceived usefulness and engagement of \tts{}. One possible explanation for this is that the lack of a physical presence of \tts{} may contribute to feelings of discomfort for some users. When interacting with a disembodied voice, people may be more likely to feel self-conscious or nervous, particularly if they have higher levels of interview anxiety. The robot's physical embodiment may have contributed to creating a more natural and intuitive interaction environment for the participants, which could have reduced their anxiety levels, hence offering them a better interview experience. However, we did not find any significant correlation between the participants' levels of anxiety and system perceptions for \sys{}. %
Further research is needed to understand underlying mechanisms and improve interview training systems' design accordingly.

\emph{Qualitative remarks. }We asked the participants their most and least favorite aspects of the systems and their suggestions for improvements as open-ended questions. The participants generally agree that the contextual feedback component common to both systems is beneficial. The design of the dialog options received criticism. Some participants found it constraining their communication during the interview. They would like to be more flexible in the question formulation instead of selecting from the predefined question sets. The participants suggested employing further cases encountered in interviews, such as when the stakeholder deviates from the themes to be covered in the interview. They would like to experience and learn from different challenging cases. Some participants proposed an interactive feedback component to receive personalized feedback based on their level of expertise as an improvement.

The participants enjoyed the emphatic gestures of the robot performed in the feedback session when using \sys{}. One participant highlighted the importance of the eye contact with the stakeholder during the interview, thanks to the physical presence of the robot. The voice quality of the robot was primarily noted as a drawback. Since the participants were non-native English speakers and the interviews are conducted in English, language impediments could have been a potential issue. The Internet connection quality may have also affected the quality of voice communication. Some participants preferred a more human-like robot capable of making facial expressions. Two of them found the robot childish and non-realistic to play the stakeholder role. Nonetheless, it is clear from the participants' remarks that they have high standards for the robotic component, and technical excellence is desired. The behavioral feedback analysis of \sys{} is also frequently noted by the participants as a significant and helpful feature. For \tts{}, mixed views regarding the voice-based agent have been expressed. One participant appreciated how it resembled the voice assistant they use every day. In contrast, another person thought it felt talking on the phone and unenjoyable. Another participant appreciated having no visual interaction during the interview, but some others complained that the agent has no visual appearance and is too artificial.

\emph{Limitations of the system. }Even though \gensys{} is engineered to accommodate any scenario prepared in the expected format, one of its current limitations is the number of available scenarios. We built an additional scenario besides the one provided by \citet{debnath2020designing}. Having a larger pool of scenarios would facilitate learners to practice the management of a broader range of requirements elicitation challenges. Interview scenarios based on prebuilt conversation graphs restrict users from expressing themselves freely. Having open-ended dialogues can help users expose themselves to diverse interview contexts. However, designing a dialogue system that can identify and address a broader spectrum of user responses poses challenges, particularly in the absence of datasets specifically created for the target task and domain~\citep{ni2023recent}. In our recent study, we investigate leveraging large language models (LLMs) for requirements elicitation interview dialogue generation through prompt engineering techniques~\citep{gorer_aire}. Our findings show that LLMs face challenges in preserving coherence and logical flow, even when dealing with a small number of interview turns. Nonetheless, these models keep a promise to perform better on task-oriented dialogue generation when pre-trained with in-domain data~\citep{gururangan2020don,chang2023survey}. 


\emph{Implications on REET. } A considerable amount of resources and effort is required to organize realistic practice requirements elicitation interviews for students. \gensys{} reduces the need for human stakeholders and can be adapted to various agent configurations and scenarios. Educators can specify their own scenarios for the desired domain, requirements, and choices for the students. Depending on the resources available, the agent who plays the stakeholder can be a humanoid robot as in \sys{}, a voice agent as in \tts{}. Other agent forms with tailored functions can also be easily utilized in the architecture. The system can be used for remote (as in our study) or in-person sessions. The participation in our study was voluntary, yet we propose including \gensys{} to the curricula of REET. The system may be used as a practice ground for students to apply their theoretical knowledge. It may also be used as a graded activity where the students are graded based on their performance. Our shared repository includes the code, scenario, and other materials needed to run our implementations and further customize the systems.

\section{Threats to Validity}
\label{sec:threats_to_validity}
In this section, we discuss the potential threats to our system and how we have attempted to mitigate them. The main threats identified, based on \citet{wohlin2012experimentation}, are as follows:

\emph{Internal validity.} Maintaining internal validity is essential to ensure that the observed effects are not due to extraneous variables but rather to the investigated treatments. In our study, we utilized a within-subject design, which involves exposing participants to multiple treatments in varying orders across participants. One common internal validity threat for within-subject design is the order effect, where the order in which treatments are administered may affect the participants' responses. To mitigate this, we used counterbalancing, a technique that involves presenting the treatments in different orders across participants. This way, the potential impact of the order of treatments is equally distributed across the participants. Moreover, to mitigate possible maturation effects, the participants experimented with the two conditions sequentially without a break. We ensured they did not receive any external training across the conditions that could impact their performance. We also adjusted the length of the interview scenario and completed the overall session within one hour to minimize possible participant fatigue across the two consecutive sessions. %

To minimize the potential social threats to internal validity, we took great care in designing the experiment's introduction for the participants. We provided only high-level context and the study's aims without disclosing details to prevent potential unequalization of experiment conditions, though every participant was treated with both systems. Participation was entirely voluntary, although the instructor encouraged the experiment as an external class exercise and rewarded it with a small bonus for the course grade. Student participants were explicitly informed that their performance during the interview would not affect their course grades. This ensured that the experiment did not generate any undue stress or pressure for the participants and that they could focus on the experiment's objectives without any external pressures.

\emph{External validity.} In our endeavor to ensure the relevance and wider applicability of our study to real-world scenarios, we carefully pursued a high degree of realism in the experimental setup. We employed realistic scenarios during the interview sessions and integrated feedback derived from actual elicitation interviews. We conducted the study within the schedule of a requirements engineering course, and the participants were recruited from this course. They were graduate students and working professionals in various positions, mostly in the software business. Their prior participation in requirements elicitation interviews and years of work experience are also diverse, as shown in Table~\ref{tab:participant_demographics}. We argue that our sample population represents the target user group, which is expected to include anyone at various levels of expertise training for requirements elicitation interviews. Still, the system should be evaluated with a broader group of users with various characteristics, including expertise, profession, as well as demographic factors such as age and nationality.

\emph{Construct validity.} Construct validity may be at risk due to participants' discomfort in feeling monitored and evaluated. This can affect the accuracy and reliability of the study's findings since the participants may not perform optimally under this feeling. To counter this social threat known as evaluation apprehension, prior to the session beginning, the participants were explicitly informed by the researcher that they would be alone during the session (see Section~\ref{sec:procedure}). The researcher was in the experiment environment for wizarding speech-to-text functionality but was kept out of sight to prevent the participants from feeling observed or judged. Our within-subject design can raise another threat to construct validity: the interaction of different treatments. When the effects of one treatment or condition interact with another, it becomes difficult to attribute the observed outcomes to a specific treatment. To address this, we used different scenarios in each condition of a participant to minimize the learning effect across the treatments. The scenarios were developed to be highly similar in terms of length and the mistakes induced but had a different context to avoid any potential learning effect across the treatments. We also examined if the scenario itself induced any confounding factor on the results and did not find any, as mentioned in Section~\ref{sec:randval}.

\emph{Conclusion validity.} %
To minimize the subjectivity, we used the well-established technology acceptance model survey to get participant opinions for the research questions interested in the users' perceptions. We also developed quantitative measures like processing speed and learning gain based on the current literature to address the related research questions objectively. To ensure that each person received the same treatment during the experiment, we established a standardized experimental technique followed consistently for all participants. The experimental systems were designed to be autonomous, except for wizarded speech-to-text functionality, thereby effectively mitigating potential researcher bias throughout the experiment. For the speech-to-text functionality, though, there is no room for the researcher to introduce any application bias as participant speech is taken as it is and accepted only if it fits one of the available dialogue options. To further improve the experiment's robustness, the technical requirements were communicated to the participants in the introduction of the experiment to eliminate the influence of any random confounding variables. We aim to mitigate any disturbing effects like background noise or distractions from others present in the experiment environment. However, since the experiment was conducted online, we were unable to exert complete control over the participants' experimental environments. Finally, all statistical tests were chosen based on the distribution and independence checks of the data to ensure that the test assumptions were not violated. %
By selecting appropriate tests and confirming the suitability of the data for each test, the statistical analyses were better able to provide reliable and meaningful results for the study. We obtain a statistical power, estimated at $1-\beta=0.68$ for the chosen alpha level $\alpha=0.05$ and an assumed effect size of $d=0.5$ (as informed in~\citet{bartlett2022have}) for the repeated measures. An increase in sample size could improve the statistical power of the results. Nevertheless, our experimental outcomes, acquired through the participation of individuals who closely resemble the target user population of the study, yield authentic and relevant findings~\citep{falessi2018empirical}.

\section{Conclusions and Future Work}
\label{sec:conclusion}
This paper introduces \gensys{}, an extensible and configurable requirements elicitation interview training system architecture to support requirements elicitation interviews training. 
We implement two instances of \gensys{}: \emph{i.} \sys{} with an embodied physical robot, and \emph{ii.} \tts{} with a voice-based virtual agent.
We assessed the two systems' advantages and drawbacks in terms of learners' experience and outcomes by conducting a user study with the students of a graduate-level REET course. %
Our research constitutes pioneering work in the field, incorporating emerging technologies to enhance the training process of requirements elicitation interviews. We share the implementations of \sys{} and \tts{} in our public repository provided in~\citet{binnur_gorer_2023_7861906}. We invite the community to further investigate and improve our work and implement other versions of \gensys{} with different agents having other interaction modalities or capabilities.  

In our study, the participants showed higher learning gains in \sys{} than in \tts{}, with a significant difference between the two systems. %
The results indicate that \tts{} was perceived to be easier to use compared to \sys{}. %
Both systems are rated appreciatively by the participants (i.e., higher than 3 = moderate level), and  we do not detect a significant difference in the attitudes and perceived usefulness between \sys{} and \tts{}. However, the participants found \tts{} to be more engaging, although objective measurements of engagement based on arousal levels of participants' speech did not indicate any significant difference between the two systems. The participants responded to both systems at similar speeds, but their responses for the turns with incorrect selection were slower than the correctly replied turns.

As part of future work, we plan to integrate one of the state-of-art speech recognition libraries~\citep{radfordrobust} into \gensys{} to replace its human-operated speech-to-text component. %
Increasing the number of available scenarios for \gensys{} would also provide additional training experience to its users. Our plan is to build a public scenario library where the RE community can contribute with their own scenarios. 
Considering the proven effect of adaptive systems in education~\citep{ahmad2017systematic}, our system can also offer a more personalized and effective learning experience by offering scenarios customized to the needs of each learner. The complexity level of scenarios and the intensity of feedback can be adapted to the learner's performance, ensuring that each learner receives personalized content appropriate for their level of expertise. In this way, experienced learners would not get bored with simple training content, and inexperienced learners would not get demotivated by overly challenging material. %

\section*{Declarations}

\paragraph{Ethics Statement}
The studies involving human participants were reviewed and approved by the Ethics Committee of Boğaziçi University. The participants provided their informed consent to participate in this study.

\paragraph{Data availability statement.} The experimental material and source code of the systems are available in the Zenodo repository, https://doi.org/10.5281/zenodo.7861906.

\paragraph{Funding.} The second author has been partially supported by the Scientific and Technological Research Council of Turkey through BIDEB 2232 grant no. 118C255. 

\paragraph{Conflicts of Interest.} The authors have no conflicts of interest to declare that are relevant to the content of this article.

\printcredits

\bibliographystyle{cas-model2-names}

\bibliography{gorer}

\bio{}
\endbio

\endbio

\end{document}